\documentclass[%
 reprint,
 amsmath,amssymb,
pra,
]{revtex4-1}

\usepackage{graphicx}
\usepackage{dcolumn}
\usepackage{bm}
\usepackage{amsmath}
\usepackage[T1]{fontenc}
\usepackage{times}
\usepackage{lmodern}
\usepackage{amssymb}
\usepackage{bm}
\usepackage{graphicx}
\usepackage{gensymb}
\usepackage{mathtools}
\usepackage{float}
\usepackage{dsfont}
\usepackage{romannum}
\usepackage{multirow}
\usepackage{amsthm}
\usepackage{colortbl}
\usepackage{subfig}
\usepackage{xr}
\usepackage{appendix}
\usepackage{textcomp,libertine}
\usepackage{hyperref}
\usepackage[mathlines]{lineno}

\usepackage[
margin=0.5in,
]{geometry}

\newlength{\hcolwidth}
\def\eR{\mathbb{R}}
\def\eC{\mathbb{C}}
\def\psib{\boldsymbol{\psi}}
\def \rhob{\boldsymbol{\rho}}

\begin{document}

%
\title{Phase retrieval for Bragg coherent diffraction imaging at high X-ray energies}
\author{S. Maddali}
\email{smaddali@anl.gov}
\author{S. O. Hruszkewycz}%
\affiliation{%
 	Materials Science Division, Argonne National Laboratory, Lemont, IL 60439 (USA)
}%

\author{M. Allain}
\affiliation{%
	Aix-Marseille Universit\'e, CNRS, Centrale Marseille, Institut Fresnel, Marseille (France)
}%

\author{W. Cha}
\author{R. Harder}
\author{J. -S. Park}
\author{P. Kenesei}
\author{J. Almer}
\affiliation{
	X-ray Sciences Division, Argonne National Laboratory, Lemont, IL 60439 (USA)
}%

\author{Y. Nashed}
\affiliation{%
	Mathematics and Computer Science Division, Argonne National Laboratory, Lemont, IL 60439 (USA)
}%

\begin{abstract}
	Coherent X-ray beams with energies $\geq 50$ keV can potentially enable three-dimensional imaging of atomic lattice distortion fields within individual crystallites in bulk polycrystalline materials through Bragg coherent diffraction imaging (BCDI).
However, the undersampling of the diffraction signal due to Fourier space compression at high X-ray energies renders conventional phase retrieval algorithms unsuitable for three-dimensional reconstruction. 
To address this problem we utilize a phase retrieval method with a Fourier constraint specifically tailored for under-sampled diffraction data measured with coarse-pitched detector pixels that bin the underlying signal. 
With our approach, we show that it is possible to reconstruct three-dimensional strained crystallites from an undersampled Bragg diffraction data set subject to pixel-area integration without having to physically upsample the diffraction signal.
Using simulations and experimental results, we demonstrate that explicit modeling of Fourier space compression during phase retrieval provides a viable means by which to invert high-energy BCDI data, which is otherwise intractable.

\end{abstract}
\maketitle

\section{Introduction}
\label{S:intro}
Coherent diffraction imaging (CDI) with radiation in the lower end of the hard X-ray range ($\sim 10$ keV) is steadily gaining traction as a technique for imaging objects ranging in size from tens of micrometers to tens of nanometers with sensitivity to diverse physical properties.
For example, when applied to Bragg reflections from single crystals, CDI and related techniques like ptychography are capable of spatial mapping of lattice imperfections like strain and crystal defects\cite{Fienup1987,Robinson2001,Robinson2009,Hruszkewycz2012,Hill2018}, thus providing a versatile tool applicable in materials science and solid state physics.
In particular, BCDI at X-ray energies $\geq 50$ keV can potentially allow the probing of nano-scale structural detail within crytalline grains of a much larger-scale bulk material. 
This is possible due to the greater penetrative power at these photon energies compared to those in present-day measurements, owing to greatly diminished absorption and extinction effects\cite{Hu2018}.
Though other high-energy X-ray methods have been developed that do not rely on beam coherence to achieve micrometer-scale resolution of grains in bulk materials (\emph{e.g.} diffraction contrast tomography and high-energy diffraction microscopy~\cite{Ludwig2008,Suter2008,Bernier2011}), implementation of high-energy BCDI has not been viable due to low coherent flux at high X-ray energies at today's synchrotron facilities.
Fortunately, improvements in synchrotron storage ring technology~\cite{Eriksson2014} now being adopted around the world will enable greatly increased coherent flux at energies greater than 50 keV, making high-energy CDI practical in the near future. 
This in turn would open up an entirely new class of possible experiments for three-dimensional high-resolution strain field mapping at such light sources. 
Specific systems of interest include individual grains embedded in bulk polycrystals subject to real-world thermo-mechanical conditions, and crystals embedded in other dense media to mimic, for example, catalytic environments.
However, in envisioning such BCDI experiments, certain difficulties can be foreseen from the standpoint of signal processing and image inversion.

Successful reconstruction of the image of a diffracting crystallite from a BCDI measurement  is predicated upon sufficient resolution of the signal features (fringe distribution about a Bragg peak). 
In any CDI experiment performed at high X-ray energies, compression of the three-dimensional Fourier space will directly impact the ability to satisfy this condition, given the fixed sizes of typical area detector pixels and practically realizable object-detector distances.
Existing methods seek to address this issue through initial signal processing by combining physical upsampling and sparsity-based methods~\cite{Chushkin2013,Maddali2018} to arrive at post-processed diffraction patterns suitable for conventional phase retrieval methods. 
In the same spirit, a recent simulation work~\cite{Pedersen2018} demonstrates the possibility of BCDI signal enhancement with refractive optical elements prior to phase retrieval.
In this article we implement a direct phase retrieval solution for undersampled BCDI data sets from compact single crystals, an approach directly applicable to future studies of embedded crystals. 
Our explicit modeling of Fourier space compression is related conceptually to an earlier work in transmission ptychography\cite{Batey2014}, in which binning-induced resolution loss due to coarse pixelation is offset by information redundancy through a high degree of probe position overlap.
Our CDI-specific work similarly focuses on phase retrieval from undersampled signals, but without incorporating signal redundancy, as in ptychography.
We precisely quantify the extent to which modeling of Fourier space compression alone allows us to relax signal sampling requirements to obtain reconstructions free of binning artifacts~\cite{Oezturk2017}.
Our work in this article (i) alludes to a fundamental theoretical limit in binning-related signal processing applications, that allows us to partially relax the well-known Nyquist criterion for CDI experiments and in addition (ii) provides an inexpensive alternative to constructing large experimental enclosures to enable sufficient angular resolution at high beam energies.
The new limit comprises a more permissive sampling criterion for successful image reconstruction, and is afforded by the additional constraints imposed on the data through modeling the compression of Fourier space during phase retrieval. 
While our focus in this article is on BCDI measurements and the reconstruction of complex three-dimensional objects, the methods described here can also be adapted to two-dimensional transmission CDI measurements.

The outline of this article is as follows: in Section~\ref{S:fourierscaling} we describe the effect of Fourier space compression on a BCDI signal and the phase retrieval framework that explicitly models this.
In Section~\ref{S:results} we present the three-dimensional reconstructions from simulated high-energy scattering as well as a reconstruction from an experimental Bragg coherent diffraction measurement designed to emulate high beam energy.
We also discuss the limits of binning for successful phase retrieval and derive the resultant sampling criterion.
In Section~\ref{S:conclusion}, we discuss the potential ramifications of this method for the design of high-energy CDI experiments.

\section{Phase retrieval with Fourier space compression}
\label{S:fourierscaling}
\begin{figure*}
	\centering
	\includegraphics[width=0.5\hcolwidth]{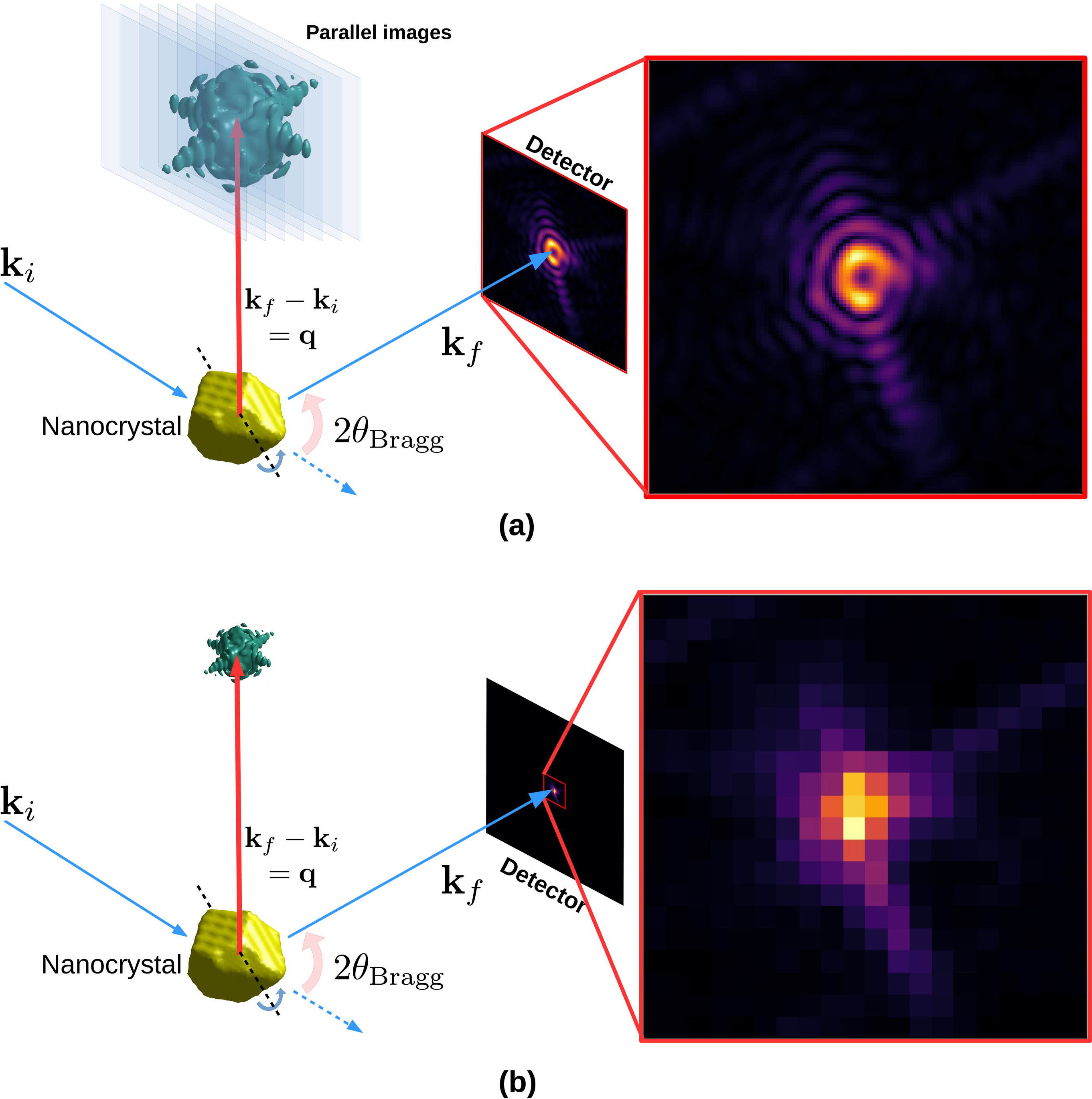}
	\caption{ 
		\textbf{(a)}
		Schematic of a conventional BCDI experiment resulting in well-sampled fringes, 
		\textbf{(b)}
		the identical experimental geometry with  photon energy 6 times higher, showing the compression of Fourier space and the signal undersampling.
		$\mathbf{k}_i$ and $\mathbf{k}_f$ are the incident and scattered wave vectors respectively and satisfy $\left|\mathbf{k}_i\right| = \left|\mathbf{k}_f\right|$. 
		In these schematics, we compare side-by-side the effects of pixel binning by considering the energy scaling factor alone, and, for the sake of simplicity, do not take into consideration the significant reduction in the Bragg angle $\theta_\text{Bragg}$ at $6$ times the original beam energy.
	}
	\label{fig:bcdi}
\end{figure*}

In a BCDI measurement, a compact single crystal coherently illuminated with monochromatic X-rays is rotated through the Bragg condition in small angular steps (typically $0.01^\circ$ steps over a $0.5^\circ$ range for a $9$ keV energy beam).
The three-dimensional scattered intensity is queried  with such a scan in a sequence of parallel slices as shown in Fig.~\ref{fig:bcdi}(a). 

\begin{figure*}
	\centering
	\includegraphics[width=0.5\hcolwidth]{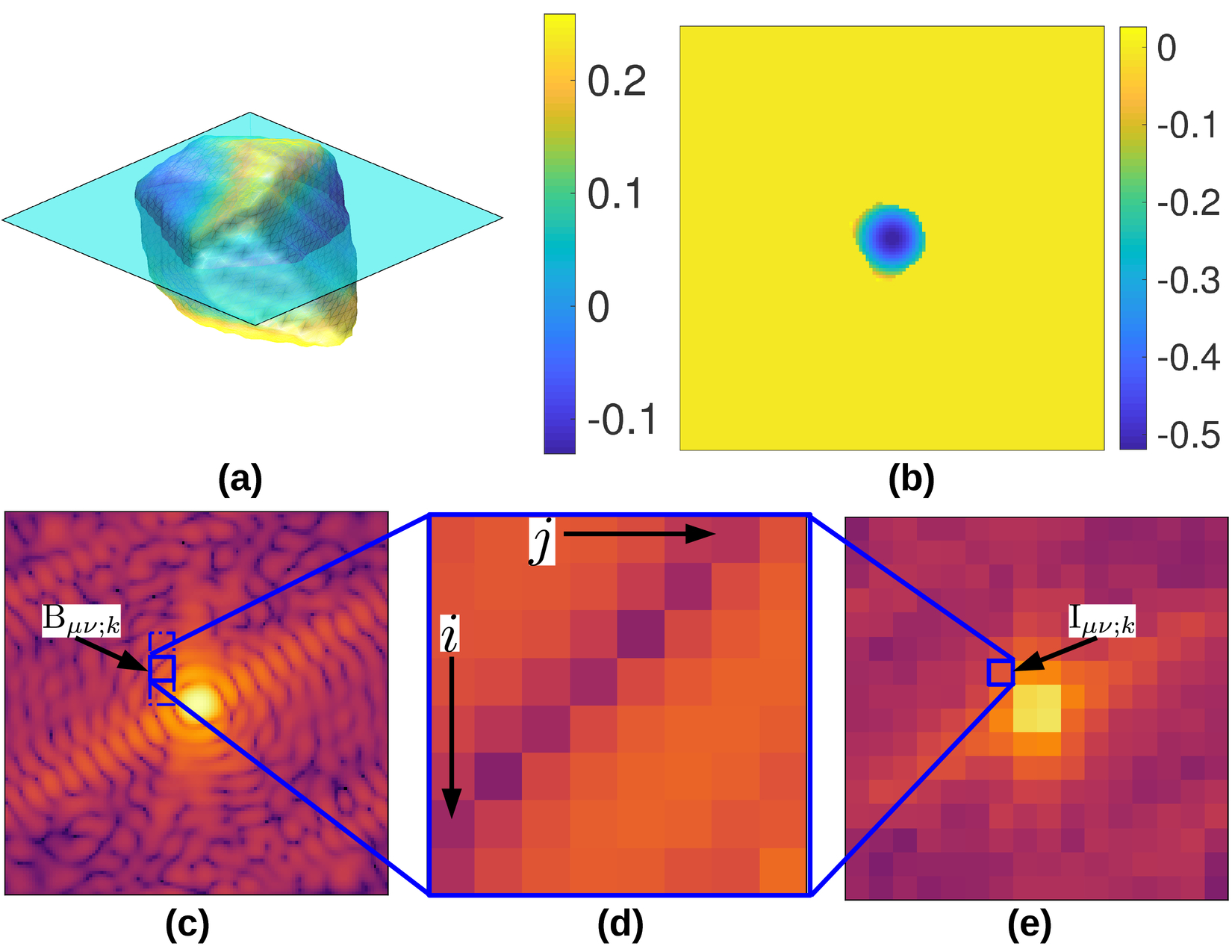}
	\caption{ 
		\textbf{(a)} A simulated complex-valued 3D object with surface phases, with the color scale in units of radians.  
		\textbf{(b)} shows a slice depicting phase through the middle of the object in (a), 
		\textbf{(c)} high-resolution diffraction pattern representing the $k$th 2D slice of the 3D intensity pattern corresponding to the object in (a). A contiguous $8\times 8$ block in this pattern is indexed by $(\mu,\nu)$, along with neighboring blocks $(\mu\pm1,\nu)$
		\textbf{(d)} zoom-in of the block in (c), 
		\textbf{(e)} binned diffraction pattern showing a pixel whose intensity is the sum of those in the block $B_{\mu\nu k}$.
	}
	\label{fig:recipspacecomp}
\end{figure*}

For a given Bragg reflection, the beam energy $E$ is inversely proportional to the Bragg angle $\theta_\text{Bragg}$.
Also since $\left|\mathbf{q}\right| = \left|\mathbf{k}_f - \mathbf{k}_i\right| \propto  2\sin\theta_\text{Bragg}$, the scale of Fourier space is inversely proportional to $E$. 
%
%
The fringe spacing and angular extents of the diffraction patterns resulting from energies $\alpha E$ and $E$ are in the ratio $1/\alpha^2$ for some multiplicative factor $\alpha$. 
Consequently, the same Fourier space aperture is resolved by proportionately fewer pixels at higher beam energies (\emph{i.e.} when $\alpha > 1$), as seen in Fig.~\ref{fig:bcdi}(b). 
The effect of this compression on the measured signal is modeled by binning the well-resolved diffraction pattern into proportionately larger pixels, and for this reason we call $\alpha$ the pixel binning factor or PBF.
Fig.~\ref{fig:recipspacecomp}(e) shows the $k$th binned intensity pattern $\mathbf{I}_k \in \eR^{M\times M}$ from a sequence of diffraction images $\mathbf{I} = \left\{\mathbf{I}_k \left|k = 1, 2, \ldots, K\right.\right\}$ acquired in a high-energy BCDI experiment. 

This signal (noisy in practice) is related to a virtual, high-resolution image $\mathbf{I}^\uparrow_k \in \eR^{N \times N}$ ( with $N > M$) that is not accessed experimentally:
\begin{equation}
	\left\langle I_{\mu \nu ; k} \right\rangle = \epsilon_{\mu \nu ; k } + \sum\limits_{i, j \in B_{\mu \nu ; k}}I^\uparrow_{ij;k}
	\label{forwardModelA}
\end{equation}
where $\left\langle \cdot \right\rangle$ denotes expectation value.
Here $I_{\mu\nu ;k}$ and $I^\uparrow_{ij;k}$ are the $(\mu,\nu)$th and $(i,j)$th pixels in $\mathbf{I}_k$ and $\mathbf{I}^\uparrow_k$ respectively, and $\epsilon_{\mu \nu ; k }$ is the contribution of the incoherent background scattering, which is generally non-zero in X-ray scattering experiments.
The index set $B_{\mu\nu;k}$ is a contiguous block of fine pixels in the virtual image $\mathbf{I}^\uparrow_k$ that subtends the same solid angle as the measured pixel $I_{\mu \nu ; k }$ (Fig.~\ref{fig:recipspacecomp}(c)).
This model intentionally does not account for compression along the third independent direction $k$ owing to the availability of high-resolution rotation stages (rotational precision $\sim 0.0003^\circ$) at coherent scattering facilities that can accommodate the compression of Fourier space along $k$ by simply reducing the angular step size of a scan.
Further, since the radius of curvature of the Ewald sphere is proportional to the beam energy, the effects of this curvature are even more diminished at $\geq 50$ keV than at typical CDI energies of $\sim8$ keV. 
For this reason we focus on the issue of detector-plane binning alone.  

The wave field $\psib_k\in\eC^{N\times N}$ associated with each $\mathbf{I}^\uparrow_k$ satisfies:
\begin{equation}
  \label{forwardModelC}
  	\mathbf{I}^\uparrow_k = |\psib_k|^2 
\end{equation}
In the Fraunhofer regime~\cite{Goodman2005}, these quantities are related to the unknown compact, three-dimensional complex-valued scatterer $\rhob$ through the discrete 3D Fourier transform operator $\mathcal{F}_{3D}$:
\begin{equation}
  \label{forwardModelD}
	\psib = \mathcal{F}_{3D}\left[\rhob\right]
\end{equation}
where $\psib\in\eC^{N\times N\times K}$ is the array built from sequential, parallel 2D slices of the diffracted field $\psib_k$.  
Our goal is to design a phase retrieval algorithm that reconstructs the well-resolved quantities $\rhob$ and $\psib$ by explicitly capturing the binning (\eqref{forwardModelA}-\eqref{forwardModelD}).
In particular, we utilize the following Fourier-space constraint (first introduced in Ref.\cite{Batey2014} in the context of ptychography) to update $\psi^{(n)}_{ij;k}$ at iteration $n$:
\begin{equation}
  \psi^{(n+1)}_{lm;k} = \left[
    \frac{ 
      I_{\mu\nu ;k}
    }{
		\epsilon_{\mu\nu;k} + \sum\limits_{i,j \in B_{\mu\nu ; k}} \left|\psi^{(n)}_{ij;k}\right|^2
    }
  \right]^\frac{1}{2} \psi^{(n)}_{lm;k} 
  \qquad
  \forall~l,m \in B_{\mu\nu ; k}
 \label{eq.fourierconstraint}
\end{equation}
This update operation ensures that the estimated 3D diffracted field $\psib^{(n)}$ at each iteration $n$ is consistent with the binning model defined in~\eqref{forwardModelA}. 
\eqref{eq.fourierconstraint} results in modified versions of the commonly used iterative algorithms in phase retrieval recipes (of which Ref.~\cite{Marchesini2007} contains a comprehensive review). 
Of these, we use the Gerchberg-Saxton error reduction (ER)\cite{Gerchberg1972} and Fienup's hybrid input-output (HIO)\cite{Fienup1987} for the reconstructions in this article.
The object support is periodically updated using a `shrinkwrap' algorithm\cite{Marchesini2003}.
%
%
%

This manner of Fourier-space signal binning is in a sense the mathematical dual process of a ptychographic fly-scan acquisition in real space, described in a recent article~\cite{Odstrcil2018}. 
In this fly-scan work, the acquired signal is modeled as the aggregate of intensity contributions from discrete probe positions in the fly-scan path. 
The resolution loss as a function of increasing scanning speed described in this fly-scan work is analogous to the case of CDI in the presence of Fourier-space compression, where we expect degradation of the final image quality from phase retrieval as a function of increased binning.
We investigate this trend comprehensively in Section~\ref{S:results}.

\eqref{eq.fourierconstraint} is also derived from a robust Gaussian model of the photon counting process~\cite{Thibault2012,Godard2012}, see Appendix~\ref{A:propt} for a detailed derivation.
This ensures that our update~\eqref{eq.fourierconstraint} is consistent with a statistically sound inference method with good asymptotic properties~\cite{Godard2012}.

Lastly, we note that at beam energies $\geq 50$ keV, there is negligible cross-talk between successive acquired images in a BCDI measurement owing to the near-orthogonality of the discrete sampling directions in Fourier space (see Appendix~\ref{A:crosstalk} for details). 
This effectively brings the high-energy versions of BCDI and transmission CDI on the same footing and allows us to describe our method in terms of the more general three-dimensional BCDI, with two-dimensional transmission CDI being a special case.
At lower beam energies, a substantial skew in the Fourier-space sampling directions results in non-negligible cross-talk between successive images and might provide BCDI with an advantage over transmission CDI in terms of the stability of such a phase retrieval algorithm.

\section{Results}
\label{S:results}
\subsection{Reconstructions from simulated scattering}
\label{SS:sim}
For simulation purposes, a synthetic complex-valued object $\rhob$ with arbitrarily oriented facets was created within a three-dimensional complex array, to represent a nanoscale crystalline particle with a definite strain state.
Pixels within the particle were assigned an amplitude of 1, and phase values that varied continuously and gradually in three dimensions. 
Pixels outside the particle were set to 0. 
The corresponding far-field scattering signal was determined from the 3D Fourier transform: $\mathbf{I}^\uparrow = \left|\mathcal{F}_{3D} \left[\rhob\right] \right|^2$.
Sufficient oversampling of the diffraction fringes in $\mathbf{I}^\uparrow$ was ensured by providing a buffer of zero-valued pixels around the particle such that the particle size was below one third of the array size in each dimension.
This ensured that the autocorrelation $\rhob \otimes \rhob$ was fully contained in the simulation array, preventing cyclic aliasing.
In our constructions, the particle and grid sizes were chosen to give a sampling rate (defined as $\sigma \equiv N/s$, where $N$ and $s$ are the pixel spans of the array and particle respectively) well above the Nyquist rate of $\sigma = 2$ in each dimension.
For various choices of the pixel binning factor $\alpha$, higher-energy diffraction patterns were simulated by binning the intensities $I^\uparrow_{ij;k}$ in the first two dimensions into pixel blocks of size $\alpha^2$ to obtain $\mathbf{I}_{k}$, after which Poisson noise was added to simulate a physical measurement. 
The values of $\alpha$ were chosen to demonstrate progressive loss of fringe visibility.
Post-binning, the pixels on the edges of the array that were not incorporated into any of the bins $B_{\mu\nu;k}$ were discarded. 
Thus, in each case the original object was recovered on a numerical grid of size $\alpha N_\alpha \times \alpha N_\alpha \times 70$, where $N_\alpha$ was the detector-plane pixel span of the binned intensity $\mathbf{I}_{k}$.
For the sake of simplicity in these simulations, the background term $\epsilon_{\mu\nu;k}$ in~\eqref{eq.fourierconstraint} was set to zero.
The signal strength was chosen to give a signal-to-noise ratio (SNR) of $\sim 40$ dB in the vicinity of the Bragg peak.
This corresponds to an approximate pixel count of $30,000$ at the Bragg peak, similar to those  in typical BCDI measurements~\cite{Yau2017,Clark2012}.
\begin{figure*}
	\centering
	\includegraphics[width=0.5\hcolwidth]{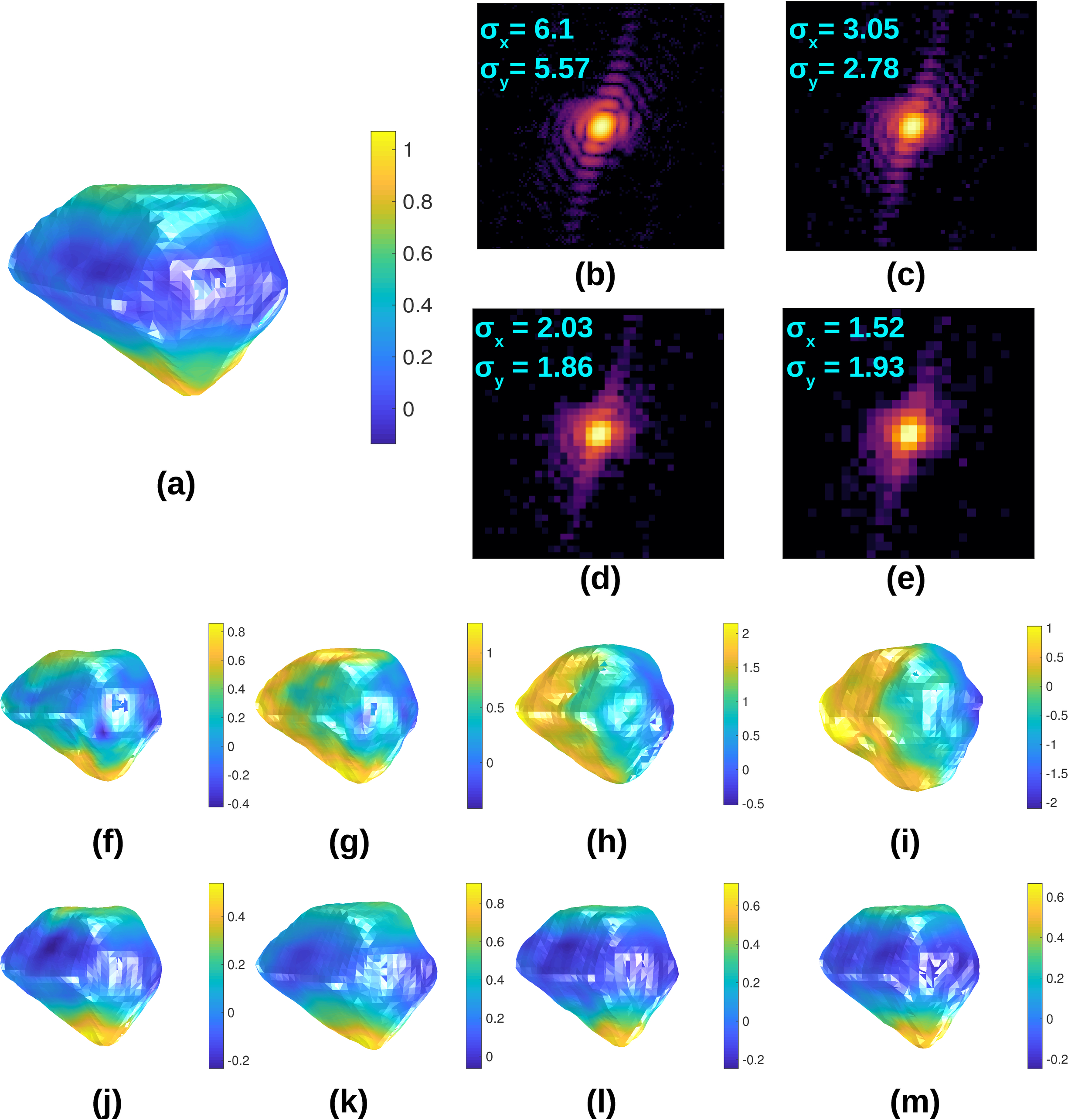}
	\caption{ 
	\textbf{(a)}: isosurface plot of a synthetic particle reconstructed with conventional phase retrieval ($\alpha = 1$), which we use as a ground truth for comparison; 
	\textbf{(b)-(e)}: central slices through binned diffraction signals, for $\alpha = 2, 4, 6, 8$ respectively, with steadily degrading fringe visibility (including sampling rates $\sigma_x, \sigma_y$ along the  axes of the imaging plane); 
	\textbf{(f)-(i)}: isosurface plots of synthetic particles recovered using conventional phase retrieval on binned noisy signals with $\alpha = 2, 4, 6, 8$; 
	\textbf{(j)-(m)}: isosurface plots of synthetic particles recovered using modified phase retrieval on binned noisy signals with $\alpha = 2, 4, 6, 8$ show much better agreement with (a). 
%
%
	}
	\label{fig:prtf_256}
\end{figure*}


A systematic comparison of modified and conventional phase retrieval schemes as a function of fringe visibility is shown in Fig.~\ref{fig:prtf_256}. 
A synthetic object of pixel span $\simeq 25$ contained in a $256\times 256 \times 70$-pixel array was used, resulting in sampling rates of $\boldsymbol{\sigma} = \left(\sigma_x, \sigma_y\right) = \left(12.19, 11.13\right)$ in orthogonal directions in the detector plane and $3.03$ in the third direction. 
Figs.~\ref{fig:prtf_256}(b)-(e) show a pronounced loss of fringe visibility as the degree of binning increases. 
As a result, the conventional phase retrieval approach that does not account for binning results in lower-quality reconstructions (Fig.~\ref{fig:prtf_256}(f)-(i)).
In contrast, accounting for Fourier space compression gives the reconstructions in Fig.~\ref{fig:prtf_256}(j)-(m), that more accurately reproduce the morphology and phase features of the reconstruction of the un-binned ($\alpha=1$) data set, which was obtained using unmodified ER and HIO algorithms (Fig.~\ref{fig:prtf_256}(a)).
Appendix~\ref{A:misc} shows phase retrieval results for several more simulated scatterers with randomly oriented facets.

We note that the binned diffraction patterns corresponding to $\alpha = 2, 4$ yielded sampling rates of $\sigma \simeq 6, 3$ respectively both of which are well above the Nyquist sampling threshold of $2$. 
In these cases, we would expect that the unmodified phase retrieval approach would be appropriate for these data and that the modified phase retrieval would not improve the image significantly.
However, in our numerical tests we see that this is not the case: the images in Fig.~\ref{fig:prtf_256}(f),(g)  (unmodified algorithm) do not reproduce the phase features of~\ref{fig:prtf_256}(a) as well as Figs.~\ref{fig:prtf_256}(j),(k), even though the morphology is reproduced faithfully. 
This is because of the inherent difference between binning and sampling in a strict signal processing sense.
Aggregation of the intensities (\emph{i.e.} binning) in each pixel block  $B_{\mu\nu;k}$ of a high-resolution diffraction pattern is a mathematical transformation that is fundamentally different from collecting a set of periodically spaced points from the high-resolution pattern (\emph{i.e.} periodic sampling), and it is only to the latter that the Nyquist criterion strictly applies.
The effects of this difference, as our tests show, are prominent at sampling rates near $2$. 
However, the binning model we have introduced can be applied in such circumstances to mitigate these artifacts.

\subsection{Limits of pixel binning}
\label{S:limits}
\begin{figure*}
	\centering
	\includegraphics[width=0.8\hcolwidth]{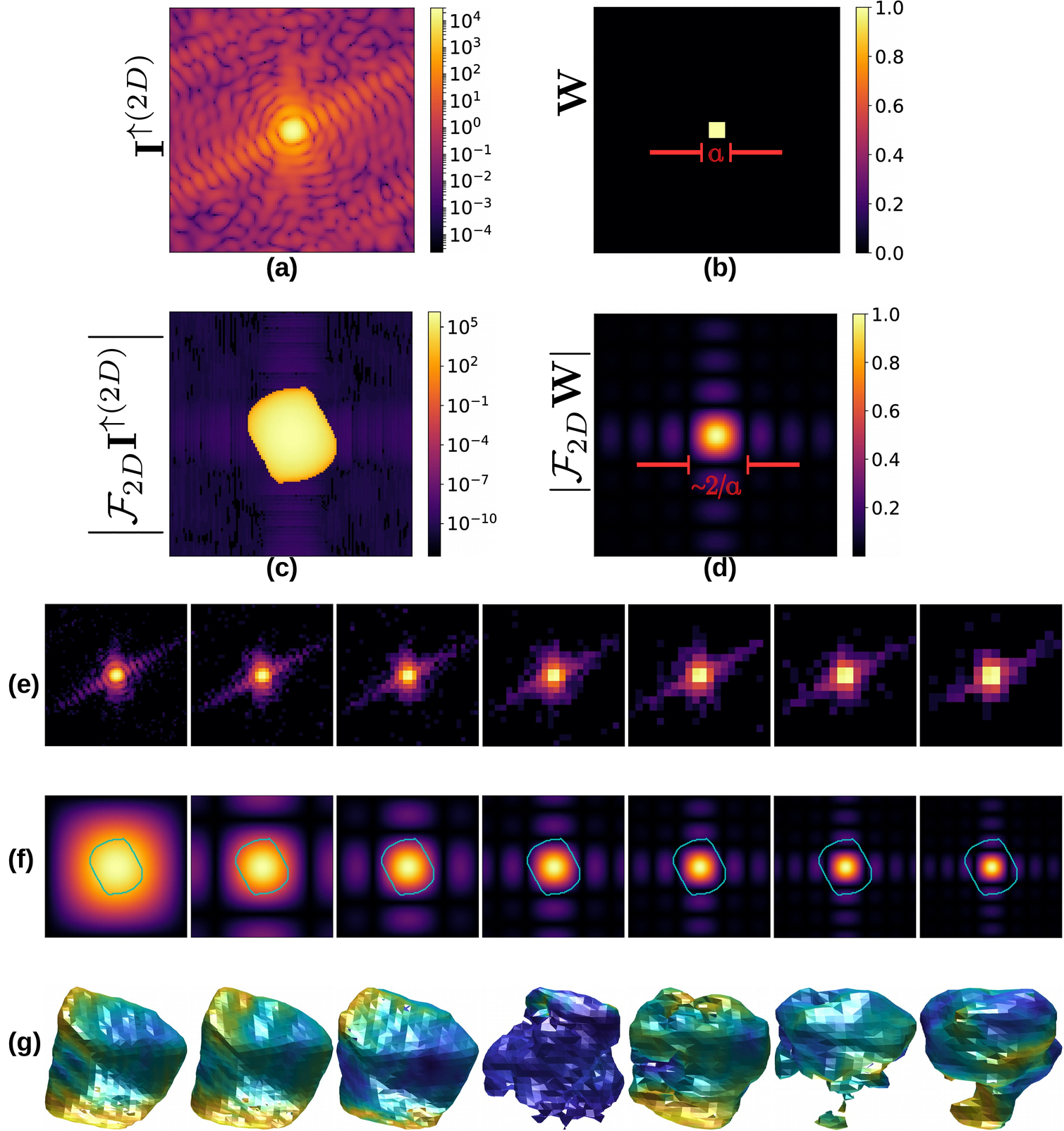}
	\caption{
		\textbf{(a)} Single detector image; 
		\textbf{(b)} Convolution kernel corresponding to binning operation; 
		\textbf{(c)} Fourier transform of the detector image (Patterson function), also equal to the object auto-correlation $\rhob \otimes \rhob$, projected in the imaging plane; 
		\textbf{(d)} 2D sinc function which is the Fourier transform of the convolution kernel in (b); 
		\textbf{(e)} Sequence of central slices of noisy signals corresponding to $\alpha = 2$ through $8$; 
		\textbf{(f)} Fourier-space representations of the corresponding convolution kernels overlaid with the outline of the Patterson function; 
		\textbf{(g)} Reconstructions from modified phase retrieval method indicating degradation of image quality from $\alpha = 5$ onwards.
	}
	\label{fig:failure}
\end{figure*}

We now address the limits of the user-defined binning (or equivalently, upsampling) parameter $\alpha$ for successful phase retrieval given a binned CDI data set. 
For the purposes of this discussion we consider this effect in terms of a single oversampled image $\mathbf{I}^{\uparrow(2D)}$ from a sequence of images (Fig.~\ref{fig:failure}(a)) without loss of generality.
We frame the analysis below by interpreting the binning operator in terms of a convolution kernel.

The pixel intensities obtained by two-dimensional binning of $\mathbf{I}^{\uparrow(2D)}$ can be thought of as periodically sampled from the 2D convolution of the oversampled intensity pattern and a 2D box function: $\mathbf{I}^{\uparrow(2D)}*\mathbf{W}$. 
The convolution kernel $\mathbf{W}$ is a square window of size $\alpha \times \alpha$ pixels whose Fourier transform is the 2D sinc function (Fig.~\ref{fig:failure}(b)). 
Further, it is known that the 2D Fourier representation of $\mathbf{I}^{\uparrow(2D)}$ is compact (see the Supplementary Material of Ref.~\cite{Maddali2018}).
We show here that the Fourier representation of $\mathbf{I}^{\uparrow(2D)} * \mathbf{W}$ (\emph{i.e.} the product of the respective 2D Fourier transforms) is indicative of the threshold of irreversible information loss.
We use $\mathcal{F}_{2D}[\cdot]$ to denote the 2D Fourier transform operator.

The sequences of images in Figs.~\ref{fig:failure}(e)-(g) correspond to increasingly large binning windows ($\alpha = 2$ through $8$), applied to a different (arbitrarily faceted) simulated nanocrystal, similar to the one featured in Fig.~\ref{fig:prtf_256}.
In the corresponding Fourier representation (Figs.~\ref{fig:failure}(f)), we see that signal information is lost when any significant component of the compact function $\mathcal{F}_{2D}\mathbf{I}^{\uparrow(2D)}$ is suppressed to zero through multiplication with $\mathcal{F}_{2D}\mathbf{W}$. 
This situation occurs at the nodes of the function $\mathcal{F}_{2D}\mathbf{W}$, for $\alpha > 4$.
Thus the criterion for successful phase retrieval is that $\mathcal{F}_{2D}\mathbf{I}^{\uparrow(2D)}$ should lie entirely within the central lobe of $\mathcal{F}_{2D}\mathbf{W}$. 
This limit is clearly demonstrated in Figure~\ref{fig:failure}(g), wherein the quality of 3D image reconstruction suffers dramatically when $\alpha$ is too high for the given particle.
An experimental ramification to this limit is that, for a fixed experimental geometry (X-ray energy, object-detector distance, detector pixel size), the largest value of $\alpha$ one can choose for phase retrieval is limited by the scatterer size. 

We now cast the above criterion in the perspective of a physical BCDI measurement.
In particular, we show that the criterion results in a relaxation by a factor of $2$ in the maximum size of the scatterer as permitted by the Nyquist condition, which is usually adhered to in CDI experiments.
Consider the relation between the Fourier-space size of a single detector pixel $\Delta q$ and the experimental parameters such as the physical pixel size $p$, sample-detector distance $z$ and the beam energy $E$: $\Delta q = Ep / hcz$, where $h$ is Planck's constant and $c$ is the speed of the propagating wave~\cite{Goodman2005}.
If $x_0$ is the largest span of the real-space scatterer, then the Nyquist criterion dictates that $\Delta q \leq 1/( 2x_0)$ for sufficient sampling of diffraction fringes and therefore successful (un-modified) phase retrieval.
This gives the criterion for the maximum permissible scatterer size for a given experimental configuration:
\begin{equation}
	x_0 \leq \left(\frac{1}{2}\right)\frac{hcz}{Ep}
	\label{eq.nyquist}
\end{equation}

The criterion derived earlier in this section, on the other hand, dictates that the span of $\mathcal{F}_{2D}\mathbf{I}^{\uparrow(2D)}$ (equal to $2x_0$ in real space units) should be no larger than the size of the central lobe of the sinc function corresponding to the single pixel window in Fourier space (equal to $2 / \Delta q$ in the same real space units).
This gives us a new scatterer size criterion:
\begin{equation}
	x_0 \leq \frac{hcz}{Ep}
	\label{eq.modifiedNyquist}
\end{equation}
which represents a factor of $2$ relaxation in the Nyquist bound of~\eqref{eq.nyquist}.
If the scatterer is larger than the threshold in~\eqref{eq.modifiedNyquist}, the modified phase retrieval method presented in this work will fail without the acquisition of additional signal data (for example, as described in Refs.~\cite{Chushkin2013,Maddali2018}), or modification of the experiment itself.
Beyond such modifications to the BCDI setup, the imaging of extended structures well beyond this size threshold by using focused high-energy X-rays (\emph{i.e.} high-energy ptychography) is currently an active area of research\cite{DaSilva2017}.

It may appear that the BCDI phase retrieval problem with Fourier space scaling is inherently underdetermined because the unknowns in the discrete quantity $\rhob \in \mathbb{C}^{N \times N \times N_2}$ outnumber the binned intensity measurements made with coarse pixels.
However, it can be shown that the unknowns and constraints are in fact equal in number:
for a fine pixel grid of size $N \times N \times N_2$, there are $2N^2 N_2$ unknowns to solve for, with the factor of $2$ arising from  the real and imaginary (or equivalently, amplitude and phase) components of $\rhob$. 
On the other hand,~\eqref{eq.fourierconstraint} imposes exactly $N^2N_2$ independent Fourier-space constraints on $\rhob$ (since for a general $\rho_{ij;k} $ we have $i,j \in \left\{1, 2, \ldots, N\right\}$ and $k \in \left\{1, 2, \ldots, N_2\right\}$) while the support constraint imposes another $N^2N_2$ independent constraints by scalar multiplication of each pixel with either $0$ or $1$. 
Even though the unknowns and constraints are equal in number, convergence of the phase retrieval to a unique solution requires that the criterion on $x_0$ above be satisfied so as to not lose information irretrievably.

In our discussion of Fig.~\ref{fig:failure}, we interpreted the binning transformation of a coherent diffraction pattern as a convolution followed by a uniform sampling operation. 
From the point of view of information loss, this convolution aspect shares an interesting parallel with the phenomenon of partial coherence in CDI.
It has been shown that the scattered intensity field in the presence of partial coherence can be treated as the convolution of the propagated coherent intensity field with a blurring kernel, typically a multivariate Gaussian function at third-generation synchrotrons~\cite{Tran2005,Clark2011,Clark2012}.
The subsequent deconvolution process to estimate the propagated wave field is computationally simpler and avoids the separate characterization of the synchrotron beam as a superposition of coherent modes. 
This modeling of partially coherent diffraction differs from the binning transformation described above only in the type of convolution kernel used, the latter being a square function as we have already seen.
Both of these processes result in reduced fringe visibility, and can be interpreted as a modulation of the auto-correlation of the diffracting object by an envelope function (which is a sinc function for binning).
In both cases, the modulating effect of the envelope can be undone to obtain the pristine signal, provided none of its significant components are suppressed by the envelope.
However, the binning case fundamentally differs from the convolution-only case  in that additional mathematical modeling is required to account for the inherent reduction in the number of measured data (\emph{i.e.} the pixel intensity counts) and provide signal sampling.

\subsection{Experimental validation}
\label{SS:exp}
\begin{figure*}
	\centering
	\includegraphics[width=0.7\hcolwidth]{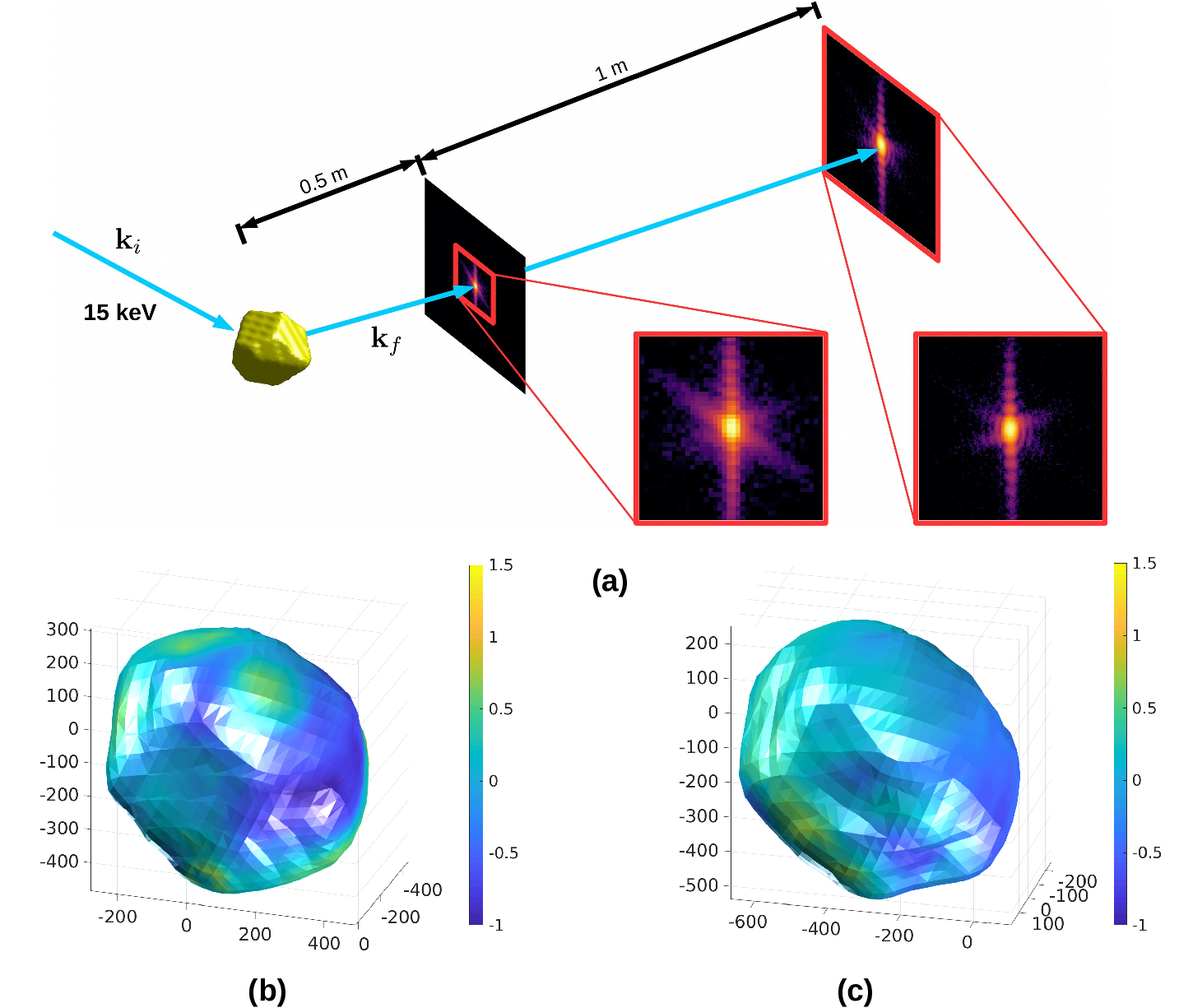}
	\caption{
		\textbf{(a)}: Experimental schematic for mock high-energy BCDI measurement; 
		\textbf{(b)}: Gold nanocrystal from modified phase retrieval acting on signal collected at a distance of $0.5$ m (isosurface colored by phase (rad)); 
		\textbf{(c)}: The same nanocrystal from conventional phase retrieval acting on signal collected at $1.5$ m. 
		The axis units in (b) and (c) are in nm.
	}
	\label{fig:binexpt}
\end{figure*}

To validate our method, we performed a synchrotron experiment in which two 3D BCDI data sets were collected from the same scatterer at the same X-ray beam energy in two different ways: (\romannum{1}) data was measured to ensure highly oversampled fringes appropriate for unmodified BCDI phase retrieval, and (\romannum{2}) data were measured under "coarse-pixel" conditions wherein the visibility of finely-spaced fringes is significantly reduced.
In our proof-of-concept experiment, the scatterer used was one of a number of gold nanocrystals of varied sizes obtained by dewetting a gold film on a silicon substrate.
The scattering measurements were made at Beamline 34-ID-C of the Advanced Photon Source. 
A BCDI data set of size $120 \times 120 \times 80$ pixels, appropriate for unmodified phase retrieval, was collected from this particle at a beam energy of $15$ keV and an object-detector distance of $1.5$ m and an angular step size of $0.005^\circ$.
The imaged particle resulting from this data set (Fig.~\ref{fig:binexpt}(b)) was found to be about $400 \times 600 \times 700$ nm in size. 
The measurement was then repeated with the same beam energy, but with the object-detector separation reduced to one third of the original distance (\emph{i.e.} $0.5$ m), resulting in significant loss of fringe visibility in the detector plane.
This configuration yields a PBF of $\alpha = 3$ relative to the first measurement because of the $3:1$ ratio of the object-detector distances.
In this sense, such a data set emulates a high-energy BCDI experiment, if one considers a factor of $3$ in energy ($45$ keV) rather than object-detector distance. 
This avoids the need to perform an actual BCDI experiment at $45$ keV with limited X-ray coherence.
Further, since the X-ray energy is the same in both cases, the image reconstructions from the two data sets correspond to identical discrete grids in real space.
The conventional phase retrieval acting on the `lower-energy-like' data set and modified phase retrieval acting on the `higher-energy-like' data set are compared in Figs.~\ref{fig:binexpt}(b) and~\ref{fig:binexpt}(c). 
The respective phase retrieval recipes are given by:
\begin{itemize}
	\item	Conventional phase retrieval: 900 iterations of ER (shrinkwrap every 30 iterations) $\rightarrow$ 600 iter. HIO $\rightarrow$ 1000 iter. ER (shrinkwrap every 50 iterations) $\rightarrow$ 3000 iter. ER with no intermediate shrinkwrap.
	\item	Modified phase retrieval: 1000 iterations of ER $\rightarrow$ 2000 iter. HIO $\rightarrow$ 1000 iter. ER $\rightarrow$ 2000 iter. HIO $\rightarrow$ 3000 iter. ER. Shrinkwrap was applied every 250 iterations during the first 2000 of ER.
Prior to this phase retrieval, the background contribution $\epsilon_{\mu\nu;k}$ to the `higher-energy' data set (assumed constant) was artifically dealt with by thresholding the pixel counts below $1$ to $0$.
\end{itemize}
In general, there is good agreement between the two reconstructions, and we speculate that the slight difference in the phase profile between Figs.~\ref{fig:binexpt}(b) and~\ref{fig:binexpt}(c) is because of differences in background levels in the two measurements. 
Namely, physical positioning of the detector closer to the direct beam in the $0.5$ m measurement increases the level of air scattering and other background sources.


We also note that because the modified Fourier-space projection of~\eqref{eq.fourierconstraint} differs from the conventional Fourier-space projection, we expect different rates of convergence, and for this reason we adopted the different phase retrieval recipes above.

\section{Conclusion}
\label{S:conclusion}
We have described a phase retrieval formalism tailored for undersampled BCDI data from high-energy X-ray scattering measurements that is based upon the modeling of signal binning due to coarse pixelation in Fourier space~\cite{Batey2014}.
The approach we describe is necessitated by the fact that at higher X-ray energies, Fourier space compression makes it impractical to resolve the fine fringe detail in a coherent diffraction pattern necessary for standard CDI phase retrieval methods.
We have demonstrated with simulations and experiments that phase retrieval algorithms explicitly designed to take into account the binning effect allows for accurate reconstruction of 3D compact crystals from data that would otherwise be intractable with standard methods. 
This is possible, to a certain limit, without acquiring any redundant data to serve as constraints, as has been done in related work~\cite{Chushkin2013,Batey2014,Maddali2018}.
Specifically, we find that our approach enables successful reconstruction of 3D images from coherent diffraction sampled up to a factor of $2$ below the Nyquist criterion in the plane of the detector.
While we have derived this limit in the context of high-energy BCDI, it could potentially be applicable to a broader class of binning-related digital imaging applications, depending on the physical origins of the signal.

Within the field of experimental X-ray and materials science, this work has important consequences for the design of space-constrained coherent scattering experiments at high-energy synchrotron facilities.
By relaxing the fringe sampling requirements and achieving the desired Fourier-space resolution through algorithmic sophistication, the relatively expensive option of building larger experimental enclosures for high-energy CDI experiments is pre-empted for a larger range of crystal sizes.
Even beamline facilities with large object-detector distances can benefit from the new ability to image larger crystals.
While enhancing the flexibility of experimental design, this is also an important step towards facilitating structural imaging experiments of nanoscale crystalline volumes in difficult-to-access environments that require the long penetration depths of high-energy X-rays.


\section{Acknowledgements}
Design and simulation of the phase retrieval framework for high-energy coherent x-ray diffraction was supported by Laboratory Directed Research and Development (LDRD) funding from Argonne National Laboratory, provided by the Director, Office of Science, of the U.S. Department of Energy under Contract No. DE-AC02-06CH11357. 
Experimental demonstration of the method was supported by the U.S. Department of Energy, Office of Science, Basic Energy Sciences, Materials Science and Engineering Division. 
This research used resources of the Advanced Photon Source, a U.S. Department of Energy (DOE) Office of Science User Facility operated for the DOE Office of Science by Argonne National Laboratory under Contract No. DE-AC02-06CH11357.

\appendix

\section{An optimization-based approach to the Fourier space constraint}
\label{A:propt}
\begin{figure}
	\centering
	\includegraphics[width=\linewidth]{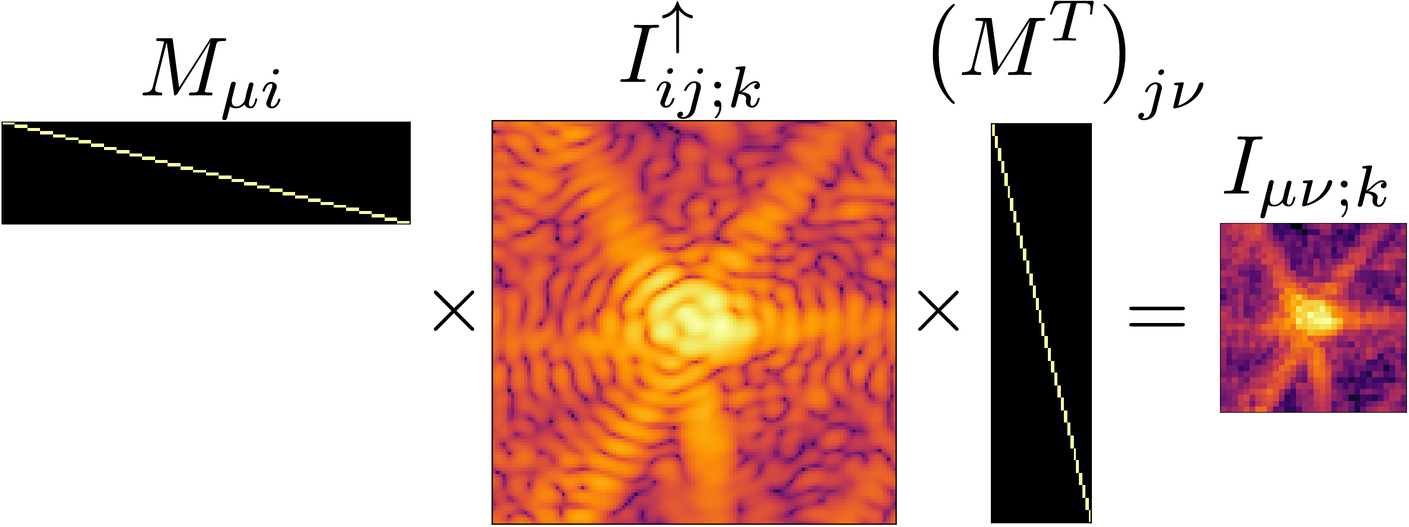}
	\caption{ 
		Visualization of the binning operation in~\eqref{eq.binningoperationmatrix} in matrix for $\alpha = 4$. 
		Here, $i, j = 1, 2, 3, \ldots, 128$ and $\mu, \nu = 1, 2, 3, \ldots, 32$.
	}
	\label{fig:binningschematic}
\end{figure}

Iterative schemes for phase retrieval alternate between updating an initial guess object between constraints in real and Fourier space. 
The Fourier space iteration typically involves comparison with some form of measured data, such as a pixelated signal on an area detector. 
The update expression for the Fourier space constraint is nonlinear and is arrived at by optimizing over one of many possible noise models associated with the measured data~\cite{Godard2012}. 
In this section we adopt this general approach to derive~\eqref{eq.fourierconstraint} more rigorously.
Given a well-sampled image of size $N \times N$ pixels from an image stack of size $N \times N \times N_2$ and a binning factor of $\alpha$, we define the binning matrix $M$ of size $(N/\alpha) \times N$ as:
\begin{equation}
	M_{\mu i} = \left\{
		\begin{matrix}
			1 & \text{if } (\mu-1) \alpha < i  \leq \mu \alpha \\
			0 & \text{otherwise}
		\end{matrix}
	\right.
	\label{eq.LR}
\end{equation}
For the $k$'th low-resolution pattern $\mathbf{I}_k$ in the sequence of diffraction images, the binning operation in~\eqref{forwardModelA} is neatly expressed as the following double summation:
\begin{align}
	\left\langle I_{\mu\nu;k} \right\rangle &= \epsilon_{\mu\nu;k} + \sum_{i=1}^{N} \sum_{j=1}^{N} M_{\mu i} \psi_{ij;k} \psi^*_{ij;k} \left(M^T\right)_{j \nu} \notag \\
	&= \epsilon_{\mu\nu;k} + \sum_i \sum_j M_{\mu i} \psi_{ij;k} \psi^*_{ij;k} M_{\nu j}
	\label{eq.binningoperationmatrix}
\end{align}
where we have used the fact that $I^\uparrow_{ij;k} = \psi_{ij;k}\psi_{ij;k}^*$.
The binning is visualized in Fig.~\ref{fig:binningschematic} as a matrix operation.

Keeping in mind that the noise associated with $I_{\mu \nu;k}$ is Poisson-dominated in a physical measurement, we focus our mathematical treatment on the variance-stabilized random variable  $\sqrt{I_{\mu \nu;k}}$ as described in Ref.~\cite{Godard2012}. 
The negative log-likelihood function for a measured dataset $I_{\mu \nu;k}$ resulting from a complex wave $\psi_{ij;k}$ in the far field is given by:
\begin{equation}
	\mathcal{L}\left[\psib\right] = \sum_{k=1}^{N_2} \sum_{\mu = 1}^{N/\alpha} \sum_{\nu=1}^{N/\alpha}
	\left[
		I_{\mu\nu;k}^{1/2}  - 
		\left\langle 
			I_{\mu\nu;k}
		\right\rangle^{1/2}
	\right]^2
	\label{eq.nll}
\end{equation}
For some $\left(i, j\right)$, the Wirtinger derivative~\cite{Sorber2012} of $\mathcal{L}$ with respect to $\psi^*_{ij;k}$ is given by:
\begin{equation}
	\delta \mathcal{L} = \sum_{\mu,\nu,k} 
	\left[
		1 - \left(
			\frac{I_{\mu\nu;k}}{\left\langle I_{\mu\nu;k} \right\rangle}
		\right)^{1/2}
	\right]
	\psi_{ij;k}
	\qquad \forall i,j 
	\label{eq.gradient}
\end{equation}
Optimality dictates that the gradient above vanishes for the wave field $\psib \equiv \hat{\psib}$ that minimizes the fitting function in~\eqref{eq.nll}. 
As a result, $\hat{\psib}$ is the solution of the following set of nonlinear equations:
\begin{equation}
	\widehat{\psi}_{ij;k} - \left( \frac{
		I_{\mu \nu ; k}
	}{
		\left\langle I_{\mu\nu;k}\right\rangle \left(\hat{\psib}\right)
	} \right)^{1/2}\widehat{\psi}_{ij;k} = 0 
	\label{eq.zerogradient}
\end{equation}
where the dependency of the expected count $\left\langle I_{\mu\nu;k}\right\rangle$ on the unknown wave field $\hat{\psib}$ is made explicit.
Starting from an initial guess $\psib^{(0)}$, the following fixed-point iteration is then usually employed to numerically solve~\eqref{eq.zerogradient}:
\begin{equation}
	\psi^{(n+1)}_{ij;k} =  \left( \frac{
		I_{\mu \nu ; k}
	}{
		\left\langle I_{\mu\nu;k}\right\rangle \left(\psib^{(n)}\right)
	} \right)^{1/2}
	\psi^{(n)}_{ij;k} \qquad \forall i, j, k
\end{equation}
This is identical to~\eqref{eq.fourierconstraint} and the original version in Ref.~\cite{Batey2014}, with the binning expressed using~\eqref{eq.binningoperationmatrix}.
While we have derived the iteration step above for an error model designed to be robust to Poisson noise, the same mathematical treatment can be applied to different models, each of which would require optimization of an objective function different from~\eqref{eq.nll}.
See Ref.~\cite{Godard2012} for details.

\section{Cross-talk between successive images in high-energy BCDI}
\label{A:crosstalk}
\begin{figure*}
	\centering
	\includegraphics[width=0.5\hcolwidth]{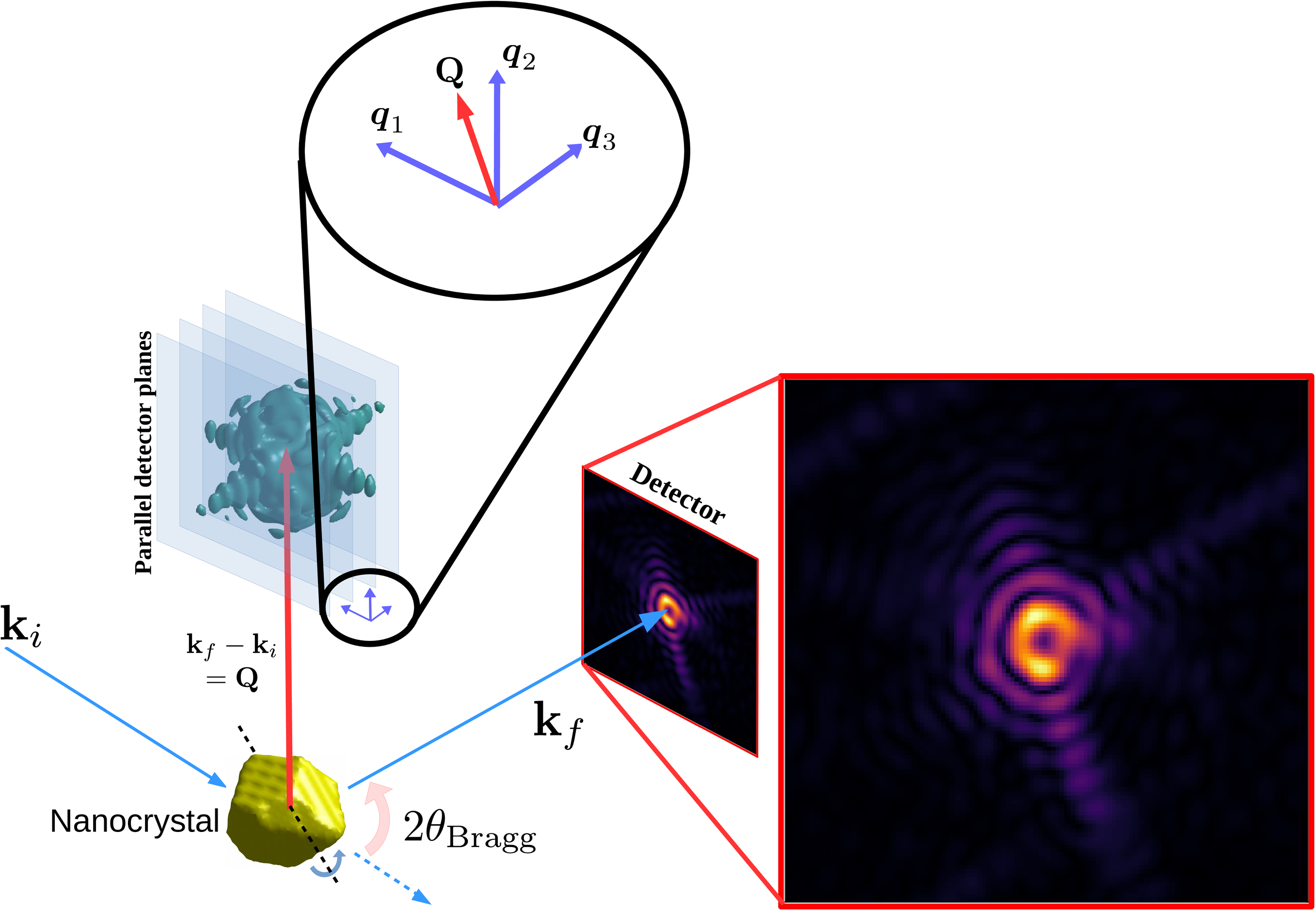}
	\caption{
		BCDI schematic showing the three vectors $\mathbf{q}_1$, $\mathbf{q}_2$ and $\mathbf{q}_3$ along which Fourier space is being sampled.
		The zoomed-in region also shows the direction of $\mathbf{Q}$ in relation to these vectors.
		Here $\mathbf{q}_1$ and $\mathbf{q}_2$ are mutually perpendicular and in the imaging plane (owing to the Cartesian pixel grid), while $\mathbf{q}_3$ is determined by the direction of change of $\mathbf{Q}$ in the rotating crystal experiment.
	}
	\label{fig:crosstalk}
\end{figure*}

Fig.~\ref{fig:crosstalk} shows the Fourier-space sampling basis vectors $\mathbf{q}_1$, $\mathbf{q}_2$, $\mathbf{q}_3$ in a BCDI measurement.
$\mathbf{q}_1$ and $\mathbf{q}_2$ are the mutually perpendicular sampling vectors in the imaging plane of the detector.
$\mathbf{q}_3$ is determined by the change in the reciprocal lattice vector $\mathbf{Q}$ on account of the the crystal rotation between two successive detector images: $\mathbf{q}_3 = -\delta\mathbf{Q}$. 
For most experimentally convenient combinations of Bragg scattering geometry, detector orientation and direction of crystal rotation, $\mathbf{q}_3$ is not perpendicular to $\mathbf{q}_1$ and $\mathbf{q}_2$, implying a non-zero projection of $\mathbf{q}_3$ into the imaging plane. 
Therefore some cross-talk usually exists between the successive images collected in a BCDI measurement.
If the detector plane is perpendicular to the exit beam, it can be shown that the angle between $\mathbf{Q}$ and the $\left\{\mathbf{q}_1, \mathbf{q}_2\right\}$ imaging plane is $\theta_\text{Bragg}$, the Bragg angle which drops to below $3^\circ$ for beam energies above $50$ keV. 
Any meaningful rotation of the crystal at this beam energy or higher would leave a very small projection of $\mathbf{q}_3$ in the imaging plane, causing minimal cross-talk and therefore minimum sharing of information between successive acquired detector images.

\section{Additional reconstruction results for various faceted scatterers}
\label{A:misc}
\begin{figure*}
	\centering
	\includegraphics[width=0.7\hcolwidth]{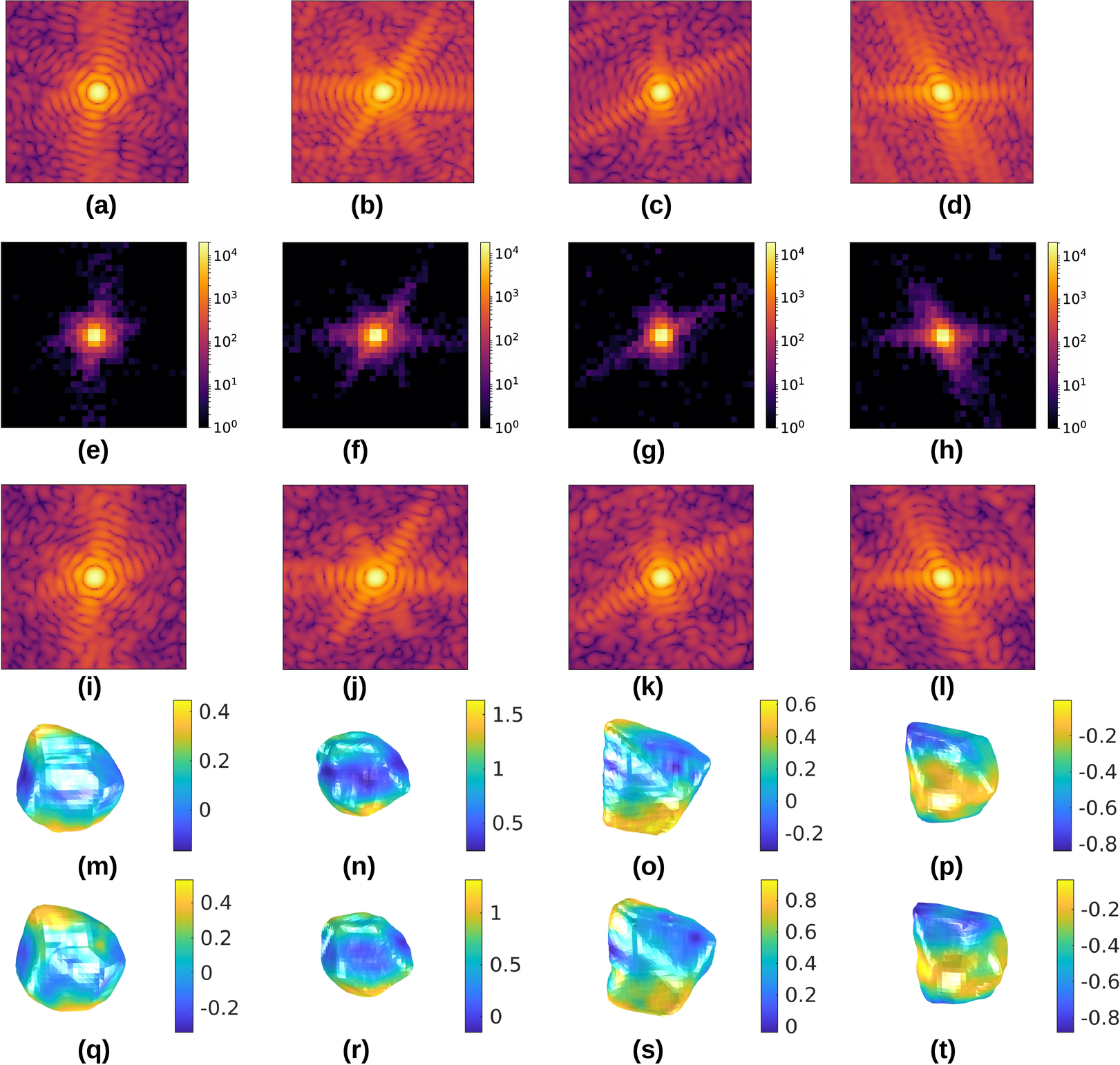}
	\caption{ 
	\textbf{(a)-(d)} Central slices through diffraction signals corresponding to 4 different simulated complex scatterers (noise-free, well-resolved fringes with sampling rate $\sigma \simeq 5.8$ in the detector plane);
	\textbf{(e)-(h)} central slices through the binned and noisy diffraction signals ($\alpha = 4$, sampling rate $\sigma \leq 1.6$); 
	\textbf{(i)-(l)} central slices through the diffraction signal recovered using the modified phase retrieval; 
	\textbf{(m)-(p)} isosurface plots of the scatterers recovered from the noisy binned signal, colored by phase (radians); 
	\textbf{(q)-(t)} isosurface plots of the scatterers recovered from the original well-resolved signal (a)-(d) with added Poisson noise.
	The conventional phase retrieval recipe to obtain images (q)-(t) was: 150 iter. ER + 100 iter. HIO + 250 iter. ER, with shrink-wrap every 50 iterations. The modified phase retrieval recipe to obtain images (m)-(p) was: 1500 iter. modified ER + 1500 iter. modified HIO + 2100 iter. modified ER, with shrink-wrap every 300 iterations.
	}
	\label{fig:variouscrystals}
\end{figure*}
Fig.~\ref{fig:variouscrystals} shows the phase retrieval results with the modified Fourier-space constraint, applied to the simulated coherent diffraction signals from four different particles, the size of each being $\simeq 22$ pixels in each dimension (smallest span was $20$ pixels) and simulated on a $128\times 128\times 70$-pixel grid.
The choice of $\alpha = 4$ for this array size resulted by design in poor fringe visibility, as we can see by computing the maximum effective sampling rate in the imaging plane: $ \sigma = 128 / ( 4 \times 20) = 1.6$, which is below the Nyquist rate of $2$.

\bibliography{MASTER}

\begin{thebibliography}{28}%
\makeatletter
\providecommand \@ifxundefined [1]{%
 \@ifx{#1\undefined}
}%
\providecommand \@ifnum [1]{%
 \ifnum #1\expandafter \@firstoftwo
 \else \expandafter \@secondoftwo
 \fi
}%
\providecommand \@ifx [1]{%
 \ifx #1\expandafter \@firstoftwo
 \else \expandafter \@secondoftwo
 \fi
}%
\providecommand \natexlab [1]{#1}%
\providecommand \enquote  [1]{``#1''}%
\providecommand \bibnamefont  [1]{#1}%
\providecommand \bibfnamefont [1]{#1}%
\providecommand \citenamefont [1]{#1}%
\providecommand \href@noop [0]{\@secondoftwo}%
\providecommand \href [0]{\begingroup \@sanitize@url \@href}%
\providecommand \@href[1]{\@@startlink{#1}\@@href}%
\providecommand \@@href[1]{\endgroup#1\@@endlink}%
\providecommand \@sanitize@url [0]{\catcode `\\12\catcode `\$12\catcode
  `\&12\catcode `\#12\catcode `\^12\catcode `\_12\catcode `\%12\relax}%
\providecommand \@@startlink[1]{}%
\providecommand \@@endlink[0]{}%
\providecommand \url  [0]{\begingroup\@sanitize@url \@url }%
\providecommand \@url [1]{\endgroup\@href {#1}{\urlprefix }}%
\providecommand \urlprefix  [0]{URL }%
\providecommand \Eprint [0]{\href }%
\providecommand \doibase [0]{http://dx.doi.org/}%
\providecommand \selectlanguage [0]{\@gobble}%
\providecommand \bibinfo  [0]{\@secondoftwo}%
\providecommand \bibfield  [0]{\@secondoftwo}%
\providecommand \translation [1]{[#1]}%
\providecommand \BibitemOpen [0]{}%
\providecommand \bibitemStop [0]{}%
\providecommand \bibitemNoStop [0]{.\EOS\space}%
\providecommand \EOS [0]{\spacefactor3000\relax}%
\providecommand \BibitemShut  [1]{\csname bibitem#1\endcsname}%
\let\auto@bib@innerbib\@empty
\bibitem [{\citenamefont {Fienup}(1987)}]{Fienup1987}%
  \BibitemOpen
  \bibfield  {author} {\bibinfo {author} {\bibfnamefont {J.~R.}\ \bibnamefont
  {Fienup}},\ }\href@noop {} {\bibfield  {journal} {\bibinfo  {journal} {JOSA
  A}\ }\textbf {\bibinfo {volume} {4}},\ \bibinfo {pages} {118} (\bibinfo
  {year} {1987})}\BibitemShut {NoStop}%
\bibitem [{\citenamefont {Robinson}\ \emph {et~al.}(2001)\citenamefont
  {Robinson}, \citenamefont {Vartanyants}, \citenamefont {Williams},
  \citenamefont {Pfeifer},\ and\ \citenamefont {Pitney}}]{Robinson2001}%
  \BibitemOpen
  \bibfield  {author} {\bibinfo {author} {\bibfnamefont {I.~K.}\ \bibnamefont
  {Robinson}}, \bibinfo {author} {\bibfnamefont {I.~A.}\ \bibnamefont
  {Vartanyants}}, \bibinfo {author} {\bibfnamefont {G.~J.}\ \bibnamefont
  {Williams}}, \bibinfo {author} {\bibfnamefont {M.~A.}\ \bibnamefont
  {Pfeifer}}, \ and\ \bibinfo {author} {\bibfnamefont {J.~A.}\ \bibnamefont
  {Pitney}},\ }\href {\doibase 10.1103/PhysRevLett.87.195505} {\bibfield
  {journal} {\bibinfo  {journal} {Phys. Rev. Lett.}\ }\textbf {\bibinfo
  {volume} {87}},\ \bibinfo {pages} {195505} (\bibinfo {year}
  {2001})}\BibitemShut {NoStop}%
\bibitem [{\citenamefont {Robinson}\ and\ \citenamefont
  {Harder}(2009)}]{Robinson2009}%
  \BibitemOpen
  \bibfield  {author} {\bibinfo {author} {\bibfnamefont {I.}~\bibnamefont
  {Robinson}}\ and\ \bibinfo {author} {\bibfnamefont {R.}~\bibnamefont
  {Harder}},\ }\href {\doibase 10.1038/nmat2400} {\bibfield  {journal}
  {\bibinfo  {journal} {Nat Mater}\ }\textbf {\bibinfo {volume} {8}},\ \bibinfo
  {pages} {291} (\bibinfo {year} {2009})}\BibitemShut {NoStop}%
\bibitem [{\citenamefont {Hruszkewycz}\ \emph {et~al.}(2012)\citenamefont
  {Hruszkewycz}, \citenamefont {Holt}, \citenamefont {Murray}, \citenamefont
  {Bruley}, \citenamefont {Holt}, \citenamefont {Tripathi}, \citenamefont
  {Shpyrko}, \citenamefont {McNulty}, \citenamefont {Highland},\ and\
  \citenamefont {Fuoss}}]{Hruszkewycz2012}%
  \BibitemOpen
  \bibfield  {author} {\bibinfo {author} {\bibfnamefont {S.}~\bibnamefont
  {Hruszkewycz}}, \bibinfo {author} {\bibfnamefont {M.}~\bibnamefont {Holt}},
  \bibinfo {author} {\bibfnamefont {C.}~\bibnamefont {Murray}}, \bibinfo
  {author} {\bibfnamefont {J.}~\bibnamefont {Bruley}}, \bibinfo {author}
  {\bibfnamefont {J.}~\bibnamefont {Holt}}, \bibinfo {author} {\bibfnamefont
  {A.}~\bibnamefont {Tripathi}}, \bibinfo {author} {\bibfnamefont
  {O.}~\bibnamefont {Shpyrko}}, \bibinfo {author} {\bibfnamefont
  {I.}~\bibnamefont {McNulty}}, \bibinfo {author} {\bibfnamefont
  {M.}~\bibnamefont {Highland}}, \ and\ \bibinfo {author} {\bibfnamefont
  {P.}~\bibnamefont {Fuoss}},\ }\href@noop {} {\bibfield  {journal} {\bibinfo
  {journal} {Nano letters}\ }\textbf {\bibinfo {volume} {12}},\ \bibinfo
  {pages} {5148} (\bibinfo {year} {2012})}\BibitemShut {NoStop}%
\bibitem [{\citenamefont {Hill}\ \emph {et~al.}(2018)\citenamefont {Hill},
  \citenamefont {Calvo-Almazan}, \citenamefont {Allain}, \citenamefont {Holt},
  \citenamefont {Ulvestad}, \citenamefont {Treu}, \citenamefont {Koblmuller},
  \citenamefont {Huang}, \citenamefont {Huang}, \citenamefont {Yan} \emph
  {et~al.}}]{Hill2018}%
  \BibitemOpen
  \bibfield  {author} {\bibinfo {author} {\bibfnamefont {M.~O.}\ \bibnamefont
  {Hill}}, \bibinfo {author} {\bibfnamefont {I.}~\bibnamefont {Calvo-Almazan}},
  \bibinfo {author} {\bibfnamefont {M.}~\bibnamefont {Allain}}, \bibinfo
  {author} {\bibfnamefont {M.~V.}\ \bibnamefont {Holt}}, \bibinfo {author}
  {\bibfnamefont {A.}~\bibnamefont {Ulvestad}}, \bibinfo {author}
  {\bibfnamefont {J.}~\bibnamefont {Treu}}, \bibinfo {author} {\bibfnamefont
  {G.}~\bibnamefont {Koblmuller}}, \bibinfo {author} {\bibfnamefont
  {C.}~\bibnamefont {Huang}}, \bibinfo {author} {\bibfnamefont
  {X.}~\bibnamefont {Huang}}, \bibinfo {author} {\bibfnamefont
  {H.}~\bibnamefont {Yan}},  \emph {et~al.},\ }\href@noop {} {\bibfield
  {journal} {\bibinfo  {journal} {Nano letters}\ }\textbf {\bibinfo {volume}
  {18}},\ \bibinfo {pages} {811} (\bibinfo {year} {2018})}\BibitemShut
  {NoStop}%
\bibitem [{\citenamefont {Hu}\ \emph {et~al.}(2018)\citenamefont {Hu},
  \citenamefont {Huang},\ and\ \citenamefont {Yan}}]{Hu2018}%
  \BibitemOpen
  \bibfield  {author} {\bibinfo {author} {\bibfnamefont {W.}~\bibnamefont
  {Hu}}, \bibinfo {author} {\bibfnamefont {X.}~\bibnamefont {Huang}}, \ and\
  \bibinfo {author} {\bibfnamefont {H.}~\bibnamefont {Yan}},\ }\href {\doibase
  10.1107/S1600576718000274} {\bibfield  {journal} {\bibinfo  {journal}
  {Journal of Applied Crystallography}\ }\textbf {\bibinfo {volume} {51}},\
  \bibinfo {pages} {167} (\bibinfo {year} {2018})}\BibitemShut {NoStop}%
\bibitem [{\citenamefont {Ludwig}\ \emph {et~al.}(2008)\citenamefont {Ludwig},
  \citenamefont {Schmidt}, \citenamefont {Lauridsen},\ and\ \citenamefont
  {Poulsen}}]{Ludwig2008}%
  \BibitemOpen
  \bibfield  {author} {\bibinfo {author} {\bibfnamefont {W.}~\bibnamefont
  {Ludwig}}, \bibinfo {author} {\bibfnamefont {S.}~\bibnamefont {Schmidt}},
  \bibinfo {author} {\bibfnamefont {E.~M.}\ \bibnamefont {Lauridsen}}, \ and\
  \bibinfo {author} {\bibfnamefont {H.~F.}\ \bibnamefont {Poulsen}},\ }\href
  {\doibase 10.1107/S0021889808001684} {\bibfield  {journal} {\bibinfo
  {journal} {Journal of Applied Crystallography}\ }\textbf {\bibinfo {volume}
  {41}},\ \bibinfo {pages} {302} (\bibinfo {year} {2008})}\BibitemShut
  {NoStop}%
\bibitem [{\citenamefont {Suter}\ \emph {et~al.}(2008)\citenamefont {Suter},
  \citenamefont {Hefferan}, \citenamefont {Li}, \citenamefont {Hennessy},
  \citenamefont {Xiao}, \citenamefont {Lienert},\ and\ \citenamefont
  {Tieman}}]{Suter2008}%
  \BibitemOpen
  \bibfield  {author} {\bibinfo {author} {\bibfnamefont {R.~M.}\ \bibnamefont
  {Suter}}, \bibinfo {author} {\bibfnamefont {C.~M.}\ \bibnamefont {Hefferan}},
  \bibinfo {author} {\bibfnamefont {S.~F.}\ \bibnamefont {Li}}, \bibinfo
  {author} {\bibfnamefont {D.}~\bibnamefont {Hennessy}}, \bibinfo {author}
  {\bibfnamefont {C.}~\bibnamefont {Xiao}}, \bibinfo {author} {\bibfnamefont
  {U.}~\bibnamefont {Lienert}}, \ and\ \bibinfo {author} {\bibfnamefont
  {B.}~\bibnamefont {Tieman}},\ }\href {\doibase 10.1115/1.2840965} {\bibfield
  {journal} {\bibinfo  {journal} {Journal of Engineering Materials and
  Technology}\ }\textbf {\bibinfo {volume} {130}},\ \bibinfo {pages} {021007}
  (\bibinfo {year} {2008})}\BibitemShut {NoStop}%
\bibitem [{\citenamefont {Bernier}\ \emph {et~al.}(2011)\citenamefont
  {Bernier}, \citenamefont {Barton}, \citenamefont {Lienert},\ and\
  \citenamefont {Miller}}]{Bernier2011}%
  \BibitemOpen
  \bibfield  {author} {\bibinfo {author} {\bibfnamefont {J.~V.}\ \bibnamefont
  {Bernier}}, \bibinfo {author} {\bibfnamefont {N.~R.}\ \bibnamefont {Barton}},
  \bibinfo {author} {\bibfnamefont {U.}~\bibnamefont {Lienert}}, \ and\
  \bibinfo {author} {\bibfnamefont {M.~P.}\ \bibnamefont {Miller}},\ }\href
  {\doibase 10.1177/0309324711405761} {\bibfield  {journal} {\bibinfo
  {journal} {The Journal of Strain Analysis for Engineering Design}\ }\textbf
  {\bibinfo {volume} {46}},\ \bibinfo {pages} {527} (\bibinfo {year} {2011})},\
  \Eprint {http://arxiv.org/abs/http://dx.doi.org/10.1177/0309324711405761}
  {http://dx.doi.org/10.1177/0309324711405761} \BibitemShut {NoStop}%
\bibitem [{\citenamefont {Eriksson}\ \emph {et~al.}(2014)\citenamefont
  {Eriksson}, \citenamefont {van~der Veen},\ and\ \citenamefont
  {Quitmann}}]{Eriksson2014}%
  \BibitemOpen
  \bibfield  {author} {\bibinfo {author} {\bibfnamefont {M.}~\bibnamefont
  {Eriksson}}, \bibinfo {author} {\bibfnamefont {J.~F.}\ \bibnamefont {van~der
  Veen}}, \ and\ \bibinfo {author} {\bibfnamefont {C.}~\bibnamefont
  {Quitmann}},\ }\href {\doibase 10.1107/S1600577514019286} {\bibfield
  {journal} {\bibinfo  {journal} {Journal of Synchrotron Radiation}\ }\textbf
  {\bibinfo {volume} {21}},\ \bibinfo {pages} {837} (\bibinfo {year}
  {2014})}\BibitemShut {NoStop}%
\bibitem [{\citenamefont {Chushkin}\ and\ \citenamefont
  {Zontone}(2013)}]{Chushkin2013}%
  \BibitemOpen
  \bibfield  {author} {\bibinfo {author} {\bibfnamefont {Y.}~\bibnamefont
  {Chushkin}}\ and\ \bibinfo {author} {\bibfnamefont {F.}~\bibnamefont
  {Zontone}},\ }\href {\doibase 10.1107/S0021889813003117} {\bibfield
  {journal} {\bibinfo  {journal} {Journal of Applied Crystallography}\ }\textbf
  {\bibinfo {volume} {46}},\ \bibinfo {pages} {319} (\bibinfo {year}
  {2013})}\BibitemShut {NoStop}%
\bibitem [{\citenamefont {Maddali}\ \emph {et~al.}(2018)\citenamefont
  {Maddali}, \citenamefont {Calvo-Almazan}, \citenamefont {Almer},
  \citenamefont {Kenesei}, \citenamefont {Park}, \citenamefont {Harder},
  \citenamefont {Nashed},\ and\ \citenamefont {Hruszkewycz}}]{Maddali2018}%
  \BibitemOpen
  \bibfield  {author} {\bibinfo {author} {\bibfnamefont {S.}~\bibnamefont
  {Maddali}}, \bibinfo {author} {\bibfnamefont {I.}~\bibnamefont
  {Calvo-Almazan}}, \bibinfo {author} {\bibfnamefont {J.}~\bibnamefont
  {Almer}}, \bibinfo {author} {\bibfnamefont {P.}~\bibnamefont {Kenesei}},
  \bibinfo {author} {\bibfnamefont {J.-S.}\ \bibnamefont {Park}}, \bibinfo
  {author} {\bibfnamefont {R.}~\bibnamefont {Harder}}, \bibinfo {author}
  {\bibfnamefont {Y.}~\bibnamefont {Nashed}}, \ and\ \bibinfo {author}
  {\bibfnamefont {S.~O.}\ \bibnamefont {Hruszkewycz}},\ }\href {\doibase
  10.1038/s41598-018-23040-y} {\bibfield  {journal} {\bibinfo  {journal}
  {Scientific Reports}\ }\textbf {\bibinfo {volume} {8}},\ \bibinfo {pages}
  {4959} (\bibinfo {year} {2018})}\BibitemShut {NoStop}%
\bibitem [{\citenamefont {Pedersen}\ \emph {et~al.}(2018)\citenamefont
  {Pedersen}, \citenamefont {Chamard},\ and\ \citenamefont
  {Poulsen}}]{Pedersen2018}%
  \BibitemOpen
  \bibfield  {author} {\bibinfo {author} {\bibfnamefont {A.~F.}\ \bibnamefont
  {Pedersen}}, \bibinfo {author} {\bibfnamefont {V.}~\bibnamefont {Chamard}}, \
  and\ \bibinfo {author} {\bibfnamefont {H.~F.}\ \bibnamefont {Poulsen}},\
  }\href@noop {} {\bibfield  {journal} {\bibinfo  {journal} {Optics express}\
  }\textbf {\bibinfo {volume} {26}},\ \bibinfo {pages} {23411} (\bibinfo {year}
  {2018})}\BibitemShut {NoStop}%
\bibitem [{\citenamefont {Batey}\ \emph {et~al.}(2014)\citenamefont {Batey},
  \citenamefont {Edo}, \citenamefont {Rau}, \citenamefont {Wagner},
  \citenamefont {Pe\ifmmode \check{s}\else \v{s}\fi{}i\ifmmode~\acute{c}\else
  \'{c}\fi{}}, \citenamefont {Waigh},\ and\ \citenamefont
  {Rodenburg}}]{Batey2014}%
  \BibitemOpen
  \bibfield  {author} {\bibinfo {author} {\bibfnamefont {D.~J.}\ \bibnamefont
  {Batey}}, \bibinfo {author} {\bibfnamefont {T.~B.}\ \bibnamefont {Edo}},
  \bibinfo {author} {\bibfnamefont {C.}~\bibnamefont {Rau}}, \bibinfo {author}
  {\bibfnamefont {U.}~\bibnamefont {Wagner}}, \bibinfo {author} {\bibfnamefont
  {Z.~D.}\ \bibnamefont {Pe\ifmmode \check{s}\else
  \v{s}\fi{}i\ifmmode~\acute{c}\else \'{c}\fi{}}}, \bibinfo {author}
  {\bibfnamefont {T.~A.}\ \bibnamefont {Waigh}}, \ and\ \bibinfo {author}
  {\bibfnamefont {J.~M.}\ \bibnamefont {Rodenburg}},\ }\href {\doibase
  10.1103/PhysRevA.89.043812} {\bibfield  {journal} {\bibinfo  {journal} {Phys.
  Rev. A}\ }\textbf {\bibinfo {volume} {89}},\ \bibinfo {pages} {043812}
  (\bibinfo {year} {2014})}\BibitemShut {NoStop}%
\bibitem [{\citenamefont {\"Ozturk}\ \emph {et~al.}(2017)\citenamefont
  {\"Ozturk}, \citenamefont {Huang}, \citenamefont {Yan}, \citenamefont
  {Robinson}, \citenamefont {Noyan},\ and\ \citenamefont {Chu}}]{Oezturk2017}%
  \BibitemOpen
  \bibfield  {author} {\bibinfo {author} {\bibfnamefont {H.}~\bibnamefont
  {\"Ozturk}}, \bibinfo {author} {\bibfnamefont {X.}~\bibnamefont {Huang}},
  \bibinfo {author} {\bibfnamefont {H.}~\bibnamefont {Yan}}, \bibinfo {author}
  {\bibfnamefont {I.~K.}\ \bibnamefont {Robinson}}, \bibinfo {author}
  {\bibfnamefont {I.~C.}\ \bibnamefont {Noyan}}, \ and\ \bibinfo {author}
  {\bibfnamefont {Y.~S.}\ \bibnamefont {Chu}},\ }\href
  {http://stacks.iop.org/1367-2630/19/i=10/a=103001} {\bibfield  {journal}
  {\bibinfo  {journal} {New Journal of Physics}\ }\textbf {\bibinfo {volume}
  {19}},\ \bibinfo {pages} {103001} (\bibinfo {year} {2017})}\BibitemShut
  {NoStop}%
\bibitem [{\citenamefont {Goodman}(2005)}]{Goodman2005}%
  \BibitemOpen
  \bibfield  {author} {\bibinfo {author} {\bibfnamefont {J.}~\bibnamefont
  {Goodman}},\ }\href {https://books.google.com/books?id=ow5xs\_Rtt9AC} {\emph
  {\bibinfo {title} {Introduction to Fourier Optics}}},\ McGraw-Hill physical
  and quantum electronics series\ (\bibinfo  {publisher} {W. H. Freeman},\
  \bibinfo {year} {2005})\BibitemShut {NoStop}%
\bibitem [{\citenamefont {Marchesini}(2007)}]{Marchesini2007}%
  \BibitemOpen
  \bibfield  {author} {\bibinfo {author} {\bibfnamefont {S.}~\bibnamefont
  {Marchesini}},\ }\href {\doibase 10.1063/1.2403783} {\bibfield  {journal}
  {\bibinfo  {journal} {Review of Scientific Instruments}\ }\textbf {\bibinfo
  {volume} {78}},\ \bibinfo {pages} {011301} (\bibinfo {year} {2007})},\
  \Eprint {http://arxiv.org/abs/http://dx.doi.org/10.1063/1.2403783}
  {http://dx.doi.org/10.1063/1.2403783} \BibitemShut {NoStop}%
\bibitem [{\citenamefont {Gerchberg}(1972)}]{Gerchberg1972}%
  \BibitemOpen
  \bibfield  {author} {\bibinfo {author} {\bibfnamefont {R.~W.}\ \bibnamefont
  {Gerchberg}},\ }\href {https://ci.nii.ac.jp/naid/10018865349/en/} {\bibfield
  {journal} {\bibinfo  {journal} {Optik}\ }\textbf {\bibinfo {volume} {35}},\
  \bibinfo {pages} {237} (\bibinfo {year} {1972})}\BibitemShut {NoStop}%
\bibitem [{\citenamefont {Marchesini}\ \emph {et~al.}(2003)\citenamefont
  {Marchesini}, \citenamefont {He}, \citenamefont {Chapman}, \citenamefont
  {Hau-Riege}, \citenamefont {Noy}, \citenamefont {Howells}, \citenamefont
  {Weierstall},\ and\ \citenamefont {Spence}}]{Marchesini2003}%
  \BibitemOpen
  \bibfield  {author} {\bibinfo {author} {\bibfnamefont {S.}~\bibnamefont
  {Marchesini}}, \bibinfo {author} {\bibfnamefont {H.}~\bibnamefont {He}},
  \bibinfo {author} {\bibfnamefont {H.~N.}\ \bibnamefont {Chapman}}, \bibinfo
  {author} {\bibfnamefont {S.~P.}\ \bibnamefont {Hau-Riege}}, \bibinfo {author}
  {\bibfnamefont {A.}~\bibnamefont {Noy}}, \bibinfo {author} {\bibfnamefont
  {M.~R.}\ \bibnamefont {Howells}}, \bibinfo {author} {\bibfnamefont
  {U.}~\bibnamefont {Weierstall}}, \ and\ \bibinfo {author} {\bibfnamefont
  {J.~C.~H.}\ \bibnamefont {Spence}},\ }\href {\doibase
  10.1103/PhysRevB.68.140101} {\bibfield  {journal} {\bibinfo  {journal} {Phys.
  Rev. B}\ }\textbf {\bibinfo {volume} {68}},\ \bibinfo {pages} {140101}
  (\bibinfo {year} {2003})}\BibitemShut {NoStop}%
\bibitem [{\citenamefont {Odstr\v{c}il}\ \emph {et~al.}(2018)\citenamefont
  {Odstr\v{c}il}, \citenamefont {Holler},\ and\ \citenamefont
  {Guizar-Sicairos}}]{Odstrcil2018}%
  \BibitemOpen
  \bibfield  {author} {\bibinfo {author} {\bibfnamefont {M.}~\bibnamefont
  {Odstr\v{c}il}}, \bibinfo {author} {\bibfnamefont {M.}~\bibnamefont
  {Holler}}, \ and\ \bibinfo {author} {\bibfnamefont {M.}~\bibnamefont
  {Guizar-Sicairos}},\ }\href {\doibase 10.1364/OE.26.012585} {\bibfield
  {journal} {\bibinfo  {journal} {Opt. Express}\ }\textbf {\bibinfo {volume}
  {26}},\ \bibinfo {pages} {12585} (\bibinfo {year} {2018})}\BibitemShut
  {NoStop}%
\bibitem [{\citenamefont {Thibault}\ and\ \citenamefont
  {Guizar-Sicairos}(2012)}]{Thibault2012}%
  \BibitemOpen
  \bibfield  {author} {\bibinfo {author} {\bibfnamefont {P.}~\bibnamefont
  {Thibault}}\ and\ \bibinfo {author} {\bibfnamefont {M.}~\bibnamefont
  {Guizar-Sicairos}},\ }\href {http://stacks.iop.org/1367-2630/14/i=6/a=063004}
  {\bibfield  {journal} {\bibinfo  {journal} {New Journal of Physics}\ }\textbf
  {\bibinfo {volume} {14}},\ \bibinfo {pages} {063004} (\bibinfo {year}
  {2012})}\BibitemShut {NoStop}%
\bibitem [{\citenamefont {Godard}\ \emph {et~al.}(2012)\citenamefont {Godard},
  \citenamefont {Allain}, \citenamefont {Chamard},\ and\ \citenamefont
  {Rodenburg}}]{Godard2012}%
  \BibitemOpen
  \bibfield  {author} {\bibinfo {author} {\bibfnamefont {P.}~\bibnamefont
  {Godard}}, \bibinfo {author} {\bibfnamefont {M.}~\bibnamefont {Allain}},
  \bibinfo {author} {\bibfnamefont {V.}~\bibnamefont {Chamard}}, \ and\
  \bibinfo {author} {\bibfnamefont {J.}~\bibnamefont {Rodenburg}},\ }\href@noop
  {} {\bibfield  {journal} {\bibinfo  {journal} {Optics express}\ }\textbf
  {\bibinfo {volume} {20}},\ \bibinfo {pages} {25914} (\bibinfo {year}
  {2012})}\BibitemShut {NoStop}%
\bibitem [{\citenamefont {Yau}\ \emph {et~al.}(2017)\citenamefont {Yau},
  \citenamefont {Cha}, \citenamefont {Kanan}, \citenamefont {Stephenson},\ and\
  \citenamefont {Ulvestad}}]{Yau2017}%
  \BibitemOpen
  \bibfield  {author} {\bibinfo {author} {\bibfnamefont {A.}~\bibnamefont
  {Yau}}, \bibinfo {author} {\bibfnamefont {W.}~\bibnamefont {Cha}}, \bibinfo
  {author} {\bibfnamefont {M.~W.}\ \bibnamefont {Kanan}}, \bibinfo {author}
  {\bibfnamefont {G.~B.}\ \bibnamefont {Stephenson}}, \ and\ \bibinfo {author}
  {\bibfnamefont {A.}~\bibnamefont {Ulvestad}},\ }\href@noop {} {\bibfield
  {journal} {\bibinfo  {journal} {Science}\ }\textbf {\bibinfo {volume}
  {356}},\ \bibinfo {pages} {739} (\bibinfo {year} {2017})}\BibitemShut
  {NoStop}%
\bibitem [{\citenamefont {Clark}\ \emph {et~al.}(2012)\citenamefont {Clark},
  \citenamefont {Huang}, \citenamefont {Harder},\ and\ \citenamefont
  {Robinson}}]{Clark2012}%
  \BibitemOpen
  \bibfield  {author} {\bibinfo {author} {\bibfnamefont {J.}~\bibnamefont
  {Clark}}, \bibinfo {author} {\bibfnamefont {X.}~\bibnamefont {Huang}},
  \bibinfo {author} {\bibfnamefont {R.}~\bibnamefont {Harder}}, \ and\ \bibinfo
  {author} {\bibfnamefont {I.}~\bibnamefont {Robinson}},\ }\href@noop {}
  {\bibfield  {journal} {\bibinfo  {journal} {Nature communications}\ }\textbf
  {\bibinfo {volume} {3}},\ \bibinfo {pages} {993} (\bibinfo {year}
  {2012})}\BibitemShut {NoStop}%
\bibitem [{\citenamefont {Da~Silva}\ \emph {et~al.}(2017)\citenamefont
  {Da~Silva}, \citenamefont {Pacureanu}, \citenamefont {Yang}, \citenamefont
  {Bohic}, \citenamefont {Morawe}, \citenamefont {Barrett},\ and\ \citenamefont
  {Cloetens}}]{DaSilva2017}%
  \BibitemOpen
  \bibfield  {author} {\bibinfo {author} {\bibfnamefont {J.~C.}\ \bibnamefont
  {Da~Silva}}, \bibinfo {author} {\bibfnamefont {A.}~\bibnamefont {Pacureanu}},
  \bibinfo {author} {\bibfnamefont {Y.}~\bibnamefont {Yang}}, \bibinfo {author}
  {\bibfnamefont {S.}~\bibnamefont {Bohic}}, \bibinfo {author} {\bibfnamefont
  {C.}~\bibnamefont {Morawe}}, \bibinfo {author} {\bibfnamefont
  {R.}~\bibnamefont {Barrett}}, \ and\ \bibinfo {author} {\bibfnamefont
  {P.}~\bibnamefont {Cloetens}},\ }\href@noop {} {\bibfield  {journal}
  {\bibinfo  {journal} {Optica}\ }\textbf {\bibinfo {volume} {4}},\ \bibinfo
  {pages} {492} (\bibinfo {year} {2017})}\BibitemShut {NoStop}%
\bibitem [{\citenamefont {Tran}\ \emph {et~al.}(2005)\citenamefont {Tran},
  \citenamefont {Peele}, \citenamefont {Roberts}, \citenamefont {Nugent},
  \citenamefont {Paterson},\ and\ \citenamefont {McNulty}}]{Tran2005}%
  \BibitemOpen
  \bibfield  {author} {\bibinfo {author} {\bibfnamefont {C.~Q.}\ \bibnamefont
  {Tran}}, \bibinfo {author} {\bibfnamefont {A.~G.}\ \bibnamefont {Peele}},
  \bibinfo {author} {\bibfnamefont {A.}~\bibnamefont {Roberts}}, \bibinfo
  {author} {\bibfnamefont {K.~A.}\ \bibnamefont {Nugent}}, \bibinfo {author}
  {\bibfnamefont {D.}~\bibnamefont {Paterson}}, \ and\ \bibinfo {author}
  {\bibfnamefont {I.}~\bibnamefont {McNulty}},\ }\href@noop {} {\bibfield
  {journal} {\bibinfo  {journal} {Optics letters}\ }\textbf {\bibinfo {volume}
  {30}},\ \bibinfo {pages} {204} (\bibinfo {year} {2005})}\BibitemShut
  {NoStop}%
\bibitem [{\citenamefont {Clark}\ and\ \citenamefont
  {Peele}(2011)}]{Clark2011}%
  \BibitemOpen
  \bibfield  {author} {\bibinfo {author} {\bibfnamefont {J.~N.}\ \bibnamefont
  {Clark}}\ and\ \bibinfo {author} {\bibfnamefont {A.~G.}\ \bibnamefont
  {Peele}},\ }\href {\doibase 10.1063/1.3650265} {\bibfield  {journal}
  {\bibinfo  {journal} {Applied Physics Letters}\ }\textbf {\bibinfo {volume}
  {99}},\ \bibinfo {pages} {154103} (\bibinfo {year} {2011})},\ \Eprint
  {http://arxiv.org/abs/https://doi.org/10.1063/1.3650265}
  {https://doi.org/10.1063/1.3650265} \BibitemShut {NoStop}%
\bibitem [{\citenamefont {Sorber}\ \emph {et~al.}(2012)\citenamefont {Sorber},
  \citenamefont {Barel},\ and\ \citenamefont {Lathauwer}}]{Sorber2012}%
  \BibitemOpen
  \bibfield  {author} {\bibinfo {author} {\bibfnamefont {L.}~\bibnamefont
  {Sorber}}, \bibinfo {author} {\bibfnamefont {M.~V.}\ \bibnamefont {Barel}}, \
  and\ \bibinfo {author} {\bibfnamefont {L.~D.}\ \bibnamefont {Lathauwer}},\
  }\href@noop {} {\bibfield  {journal} {\bibinfo  {journal} {SIAM Journal on
  Optimization}\ }\textbf {\bibinfo {volume} {22}},\ \bibinfo {pages} {879}
  (\bibinfo {year} {2012})}\BibitemShut {NoStop}%
\end{thebibliography}%
\end{document}